%% file: prd_wh1invfb.tex
\documentclass[aps,prd,preprint,superscriptaddress,showpacs,floatfix,nofootinbib]{revtex4}
\usepackage{graphicx}
\usepackage{bm}

\newcommand{\MET}{\mbox{$\not\!\!E_T$}}

\begin{document}

\preprint{FERMILAB-PUB-08-070-E}

\title{Search for Standard Model Higgs Boson Production in Association
 with a $W$ Boson at CDF}

\input{Jan_2008_Authors_Visitors1}


\date{\today}

\begin{abstract}
We present a search for standard model Higgs boson production in
association with a $W$ boson in proton-antiproton collisions
($p\bar{p}\rightarrow W^\pm H \rightarrow \ell\nu b\bar{b}$) at a
center of mass energy of 1.96 TeV. The search employs data collected
with the CDF II detector which correspond to an integrated luminosity
of approximately 1 fb$^{-1}$.  We select events consistent with a
signature of a single lepton ($e^\pm/\mu^\pm$), missing transverse
energy, and two jets. Jets corresponding to bottom quarks are
identified with a secondary vertex tagging method and a neural network
filter technique. The observed number of events and the dijet mass
distributions are consistent with the standard model background
expectations, and we set 95\% confidence level upper limits on the
production cross section times branching ratio ranging from 3.9 to 1.3~pb for
Higgs boson masses from 110 to $150\,\mathrm{GeV}/c^2$, respectively.
\end{abstract}

\pacs{13.85.Rm, 14.80.Bn}

\maketitle

\section{Introduction}

Standard electroweak theory predicts a single fundamental scalar
particle, the Higgs boson, which arises as a result of spontaneous
electroweak symmetry breaking~\cite{Higgs:1964pj}; however, the Higgs
boson has not been direct observed experimentally.  The current
constraint on the Higgs boson mass, $m_H > 114.4\,\mathrm{GeV}/c^2$ at
95\% confidence level (C.L.), comes from direct searches
at LEP2 experiments~\cite{Barate:2003sz}.  Global fits to electroweak
measurements exclude masses above $144\,\mathrm{GeV}/c^2$ at 95\%
CL~\cite{Alcaraz:2006mx}.

At the Tevatron $p\bar p$ collider at Fermilab, the
next-to-leading-order (NLO) Higgs boson production cross section by
gluon fusion is about ten times larger than for $WH$ associated
production, and the cross section for $WH$ is about twice that of
$ZH$~\cite{Han:1991ia}.  The Higgs boson decay branching ratio is
dominated by $H\rightarrow b\bar{b}$ for $m_H<135\,\mathrm{GeV}/c^2$
and by $H\rightarrow W^+W^-$ for
$m_H>135\,\mathrm{GeV}/c^2$~\cite{Djouadi:1997yw}.  Background QCD
$b\bar b$ production processes in the same invariant mass range have
cross sections at least four orders of magnitude greater than that of
Higgs boson production~\cite{Abulencia:2006ps}, and this renders
searches in the $gg\rightarrow H
\rightarrow b\bar{b}$ channel extremely difficult.  However, requiring the
leptonic decay of the associated weak boson reduces the huge QCD
background rate. As a result, $WH\rightarrow \ell
\nu b\bar{b}$ is considered to be one of the most sensitive processes
for low mass Higgs boson searches~\footnote{In this paper, lepton
($\ell$) denotes electron ($e^\pm$) or muon ($\mu^\pm$), and neutrino
($\nu$) denotes electron neutrino ($e_\nu$) or muon neutrino
($\mu_\nu$).}.

Searches for $WH\rightarrow \ell\nu b\bar{b}$ at
$\sqrt{s}=1.96\,\rm{TeV}$ have been most recently reported by CDF
(using data corresponding to an integrated luminosity of
319~pb$^{-1}$)\cite{Abulencia:2006aj} and D0
(440~pb$^{-1}$)\cite{new_d0}.  The CDF analysis used a secondary
vertex $b$-tagging algorithm ({\sc secvtx}) to distinguish $b$-quark
jets from light flavor or gluon jets~\cite{Acosta:2004hw}.  Upper
limits on the Higgs boson production rate, defined as the cross
section times branching ratio ($\sigma \cdot {\cal B}$), were derived
for mass hypotheses ranging from 110 to $150\,\mathrm{GeV}/c^2$.  The
rate was constrained to be less than 10~pb at 95\% C.L. for $m_H=110$
and less than 2.8~pb for $150\,\mathrm{GeV}/c^2$.  In that analysis,
about 50\% of the jets tagged by the {\sc secvtx} tagging algorithm
were actually falsely b-tagged jets originating from light flavor,
gluon, or charm quarks.  This effect is due to the finite resolution
of track measurements and the long lifetime of $D$ mesons. Even the
small fraction of mistagged events in the dominant $Wq\bar q$ process
is significant compared to true $Wb\bar b$ production.  To reduce this
contamination and enhance the $b$-jet purity of our sample, we
introduce a $b$-tagging neural network filter which uses as inputs jet
characteristics as well as secondary vertex information.

In this paper, we present a search for $WH\rightarrow
\ell\nu b\bar{b}$ production at CDF using about 1 fb$^{-1}$ of data.
Section~\ref{sec:detector} describes the CDF II
detector. The event selection criteria are explained in Sec.~\ref{sec:eventSelection}.
In Sec.~\ref{sec:btagging}, the $b$-tagging
algorithm with {\sc secvtx} and neural network (NN) are discussed in
detail.  Contributions from the standard model (SM) background are
calculated in Sec.~\ref{sec:bkg} for various sources. In
Sec.~\ref{sec:Acceptance}, signal acceptance and systematic
uncertainties are estimated.  The search optimization and
statistical interpretation of the results are presented in
Secs.~\ref{sec:interpretation} and~\ref{sec:limit}, respectively.
Finally, our conclusions are presented in Sec.~\ref{sec:conclusions}.

\section{CDF II Detector}
\label{sec:detector}
The CDF II detector geometry is described using a cylindrical
coordinate system~\cite{Acosta:2004yw}.  The $z$-axis follows the
proton direction, and the polar angle $\theta$ is usually expressed
through the pseudorapidity $\eta = -\ln(\tan(\theta/2))$.  The
detector is approximately symmetric in $\eta$ and in the azimuthal
angle~$\phi$.  

Charged particles are tracked by a system of silicon microstrip
detectors and a large open cell drift chamber in the region
$|\eta|\leq 2.0$ and $|\eta|\leq 1.0$, respectively.  The tracking
detectors are immersed in a $1.4\,\mathrm{T}$ solenoidal magnetic
field aligned coaxially with the incoming beams, allowing measurement
of charged particle momentum transverse to the beamline.

The resolution on the transverse momentum $p_T = p \sin \theta$ is
measured to be $\delta p_T/p_T
\approx 0.1\% \cdot p_T$(GeV) for the combined tracking system.
The resolution on the track impact parameter ($d_0$), or distance from
the beamline axis to the track at the track's closest approach in the
transverse plane, is $\sigma(d_0) \approx 40\,\mu{\rm m}$, about
$30\,\mu{\rm m}$ of which is due to the transverse size of the
Tevatron interaction region.

Outside of the tracking systems and the solenoid, segmented
calorimeters with projective tower geometry are used to reconstruct
electromagnetic showers and hadronic
jets~\cite{Balka:1987ty,Bertolucci:1987zn,Albrow:2001jw} over the
pseudo-rapidity range $|\eta|<3.6$.  A transverse energy $E_T =
E\sin{\theta}$ is measured in each calorimeter tower where the polar
angle ($\theta$) is calculated using the measured $z$ position of the event
vertex and the tower location.

Small contiguous groups of calorimeter towers with signals are
identified and summed together into an energy cluster.  Electron
candidates are identified in the central electromagnetic calorimeter
(CEM) as isolated, mostly electromagnetic clusters which match a track
in the pseudorapidity range $|\eta|<1.1$.  The electron transverse
energy is reconstructed from the electromagnetic cluster with a
resolution $\sigma(E_T)/E_T = 13.5\%/\sqrt{E_T/(\mathrm{GeV})} \oplus
2\%$~\cite{Balka:1987ty}.  Jets are identified as a group of
electromagnetic (EM) and hadronic (HAD) calorimeter clusters which
fall within a cone of radius $\Delta{R}=\sqrt{\Delta
\phi^2 + \Delta \eta^2} \leq 0.4$ units around a high-$E_T$ seed
cluster~\cite{Abe:1991ui}.  Jet energies are corrected for calorimeter
non-linearity, losses in the gaps between towers, multiple primary
interactions, out-of-cone losses, and inflow from underlying
event~\cite{Bhatti:2005ai}. 

For this analysis, muons are detected in three separate subdetectors.
After at least five interaction lengths in the calorimeter, the muons
first encounter four layers of planar drift chambers (CMU), capable of
detecting muons with $p_T > 1.4\,\rm{GeV}/c$~\cite{Ascoli:1987av}.
Four additional layers of planar drift chambers (CMP) behind another
60~cm of steel detect muons with $p_T > 2.8$
GeV/c~\cite{Dorigo:2000ip}.  These two systems cover the same central
pseudorapidity region with $|\eta| \leq 0.6$.  Muons which exit the
calorimeters at $ 0.6 \leq |\eta| \leq 1.0$ are tracked by the CMX
detector, consisting of four layers of drift chambers.  Muon
candidates are then identified as isolated tracks which extrapolate to
line segments or ``stubs'' in one of the muon subdetectors.  A track
which is linked to both CMU and CMP stubs is called a CMUP muon.

The CDF trigger system is a three-level filter, with tracking
information available even at the first level~\cite{Thomson:2002xp}.
Events used in this analysis have all passed the high-energy electron
or muon trigger selection.  The first stage of the central electron
trigger requires a track with $p_T > 8$~GeV/c pointing to a tower with
$E_T > 8$~GeV and $E_{\mathrm{HAD}}/E_{\mathrm{EM}}<0.125$.  The first
stage of the muon trigger requires a track with $p_T > 4$~GeV/$c$
(CMUP) or 8~GeV/$c$ (CMX) pointing to a muon stub.  A complete lepton
reconstruction is performed online in the final trigger stage, where
we require $E_T > 18\,\mathrm{GeV}/c^2$ for electrons and $p_T >
18\,\mathrm{GeV}/c$ for muons.

\section{Event Selection}
\label{sec:eventSelection}
The observable final state from the $WH\rightarrow \ell\nu b\bar b$
signal consists of two jets plus a lepton and missing transverse
energy. The leptonic $W$ decay requirement in $WH$ events yields the
high-$p_T$ lepton and large missing transverse energy due to the
neutrino.

The results presented here use data collected between February 2002
and February 2006.  The data collected using the CEM and CMUP
triggers correspond to $955 \pm 57$~pb$^{-1}$, while the data from the
CMX trigger corresponds to $941 \pm 56$~pb$^{-1}$.  

The missing transverse energy (\MET)\ is a reconstructed quantity
that is
defined as the opposite of the vector sum of all calorimeter tower
energy depositions projected on the transverse plane.  It is often
used as a measure of the sum of the transverse momenta of the
particles that escape detection, most notably neutrinos.  To be more
readily interpretable as such, the raw
\MET\ vector is adjusted for corrected jet energies, for the transverse momentum of the muons, and for
the energy deposition of any minimum ionizing high-$p_T$ muons.

Events are considered as $WH$ candidates only if they have exactly one
high-$p_T$ isolated lepton~\cite{Acosta:2004uq}, with
$E_T>20\,\mathrm{GeV}$ for electrons or $p_T>20\,\mathrm{GeV}/c$ for
muons.  The isolation cone of $\Delta R=0.4$ surrounding the lepton
must have less than 10\% of the lepton energy.  A primary event vertex
position is calculated by fitting a subset of particle tracks which
are consistent with having come from the beamline.  The distance
between this primary event vertex and the lepton track $z_0$ must be
less than 5~cm to ensure the lepton and the jets come from
the same hard interaction.  Some leptonic $Z$ decays would mimic the
single-lepton signature if a lepton is unidentified.  Events
are therefore rejected if a second track with $p_T>10\,\mathrm{GeV}/c$
forms an invariant mass with the lepton which falls in the $Z$-boson
mass window ($76< m_{\ell X} < 106\,\mathrm{GeV}/c^2$).  The selected events are
required to have \MET\ greater than 20 GeV.

The $WH$ signal includes two jets originating from $H\rightarrow
b\bar{b}$ decays; these jets are expected to have large transverse
energy.  The jets are required to be in the pseudorapidity range
covered by the silicon detector so that secondary vertices from $b$
decays can be reconstructed.  Specifically, we require the jets
satisfy $E_T>15$~GeV and $|\eta|<2.0$.  The search for $WH\rightarrow
\ell\nu b\bar b$ is performed in the sample of events with $W$+
exactly 2 jets; however, samples of events with $W$+1,3,$\geq$4 jets
are used to cross-check the background modeling.
   
To increase the signal purity of the $W$+2-jet events, at least one
jet must be $b$-tagged by the {\sc secvtx} algorithm.  If only one of the
jets is $b$-tagged, the jet must also pass the NN
$b$-tagging filter.  If there are two or more {\sc secvtx} $b$-tagged jets,
the NN is not applied.
With a {\sc secvtx} mistag rate of 1\%, it is rare that two or more
jets in the same events are mistagged by {\sc secvtx}.

\section{Secondary vertex $b$-Tagging}
\label{sec:btagging}

Multijet final states have dominant contributions from QCD light
flavor jet production, but the standard model Higgs boson decays
predominantly to bottom quark pairs. Correctly identifying the $b$
quark jets helps to remove most of the QCD background.  An algorithm
has been developed and used to tag displaced secondary vertices from
$b$ quark decays; however, the sample tagged by the {\sc secvtx}
algorithm still has significant contamination from falsely-tagged
light-flavor or gluon jets and the misidentification of $c$ quarks as
$b$-jets~\cite{Sal:1}. This search introduces a multivariate NN
technique intended to improve the {\sc secvtx} tagging purity.

The $b$-quark has a relatively long lifetime,
and $B$ hadrons formed during the hadronization of the initial $b$
quark can travel a significant distance on the order of millimeters
before decaying into a collection of lighter hadrons.  The decay
vertex can be reconstructed by identifying tracks which form a
secondary vertex significantly displaced from the $p\bar{p}$
interaction point (primary vertex).

The {\sc secvtx} $b$-tagging algorithm is applied to each jet in the
event, using only the tracks which are within $\eta$-$\phi$ distance
of $\Delta R=0.4$ of the jet direction.  Displaced tracks in jets are
used for the {\sc secvtx} reconstruction and are distinguished by a
large impact parameter significance ($|d_0/\sigma_{d_0}|$) where $d_0$
and $\sigma_{d_0}$ are the impact parameter and the total uncertainty
from tracking and beam position measurements.  Secondary vertices are
reconstructed with a two-pass approach which tests for high-quality vertices 
in the first pass and allows lower-quality vertices in the second pass.
In pass 1, at least three tracks are required to pass loose selection criteria
($p_T>0.5\,\mathrm{GeV}/c$, $|d_0/\sigma_{d_0}|>2.0$), and a secondary vertex is
fit from the selected tracks.  One of the tracks used in the
reconstruction is required to have $p_T>1.0\,\mathrm{GeV}/c$.  If pass 1 fails,
then a vertex is sought in pass 2 from at least two tracks satisfying
tight selection criteria ($p_T>1.0\,\mathrm{GeV}/c$, $|d_0/\sigma_{d_0}|>3.5$
and one of the pass 2 tracks must have $p_T>1.5\,\mathrm{GeV}/c$).  If either
pass is successful, the transverse distance ($L_{xy}$) from the primary
vertex of the event is calculated along with the associated
uncertainty.  This uncertainty $\sigma_{L_{xy}}$ includes
the uncertainty on the primary vertex position.  Finally jets are
tagged positively or negatively depending on the $L_{xy}$ significance
($L_{xy}/\sigma_{L_{xy}}$):
\begin{eqnarray} L_{xy}/\sigma_{ L_{xy}} &\geq& 7.5\ \ \quad
{\mathrm{(positive\ tag)}} \label{eq:postag} \\ L_{xy}/\sigma_{
L_{xy}} &\leq& -7.5 \quad {\mathrm{(negative\ tag)}} \label{eq:negtag}
\end{eqnarray}
These values have been tuned for optimum efficiency and purity in
simulated $b$-jet samples from decays of top quarks.  The energy
spectrum for those jets is similar to the spectrum for $b$ jets from
decays of Higgs bosons.

The sign of $L_{xy}$ indicates the position of the secondary vertex
with respect to the primary vertex along the direction of the jet.  If
the angle between the jet axis and the vector pointing from the primary
vertex to the secondary vertex is less than $\pi/2$, $L_{xy}$ is
positively defined; otherwise, it is negative. If $L_{xy}$ is
positive, the secondary vertex points towards the direction of the
jet, as in true $B$ hadron decays.  For negative $L_{xy}$ the
secondary vertex points away from the jet; this may happen as a result
of mismeasured tracks, so jets tagged with a negative $L_{xy}$ are
labeled mistagged jets.  In order to reject secondary vertices due to
material interaction, the algorithm vetoes two-track vertices found
between 1.2 and 1.5 cm from the center of the silicon detector (the
inner radius of the beampipe and the outer radius of the innermost
silicon layer being within this range).  All vertices more than 2.5 cm
from the center are rejected.

The negative tags are useful for evaluating the rate of false positive
tags, which are defined ``mistags'' in the background estimates.
Mismeasurements are expected to occur randomly; therefore the $L_{xy}$
distribution of fake tags is expected to be symmetric with respect to
zero.  Simulated events are used to correct a small asymmetry due to
true long-lived particles in light flavor jets.

The efficiency for identifying a secondary vertex is
found to be different in the simulated and observed datasets.  We measure an
efficiency scale factor, which is defined as the ratio of the observed to the simulated
efficiencies, to be $0.91\pm0.06$ in a sample of high-$E_T$ jets
enriched in $b$ jets by requiring a soft lepton ($p_T >
8\,\mathrm{GeV}/c^2$) from semileptonic heavy quark
decays~\cite{Acosta:2004hw}.

Secondary vertex {\sc secvtx} $b$-tagging exploits the long lifetime
of $B$ hadrons. $D$ hadrons originating from $c$-quarks also have
fairly long lifetime, and secondary vertices in $c$-jets are
frequently tagged.  Therefore jets tagged by {\sc secvtx} are
contaminated not only by falsely tagged light flavor ($uds$ or gluon)
jets, but also by long-lived charmed hadrons in $c$-jets.
A neural network has
been developed to filter the $b$-tagging results in order to improve
the $b$-tagging purity.

The neural network used in this article employs the {\sc
jetnet}\cite{Peterson:1993nk} package. The tagger is designed with two
networks in series. The $b-l$ network is trained to separate
$b$-jets from light-quark jets ($l$-jets), and the $b-c$ network is
trained to separate $b$-jets from $c$-jets. Jets which pass a cut on
both of the NN outputs are accepted by the tagger. These
neural networks are trained and applied only to events which are
already tagged by the {\sc secvtx} algorithm.  The current NN
$b$-tagging is tuned to increase the purity of the {\sc secvtx}
$b$-tagged jets, not to increase the tagging efficiency.

\begin{figure}[htbp]
 \begin{center}
  \includegraphics[width=0.48\textwidth]{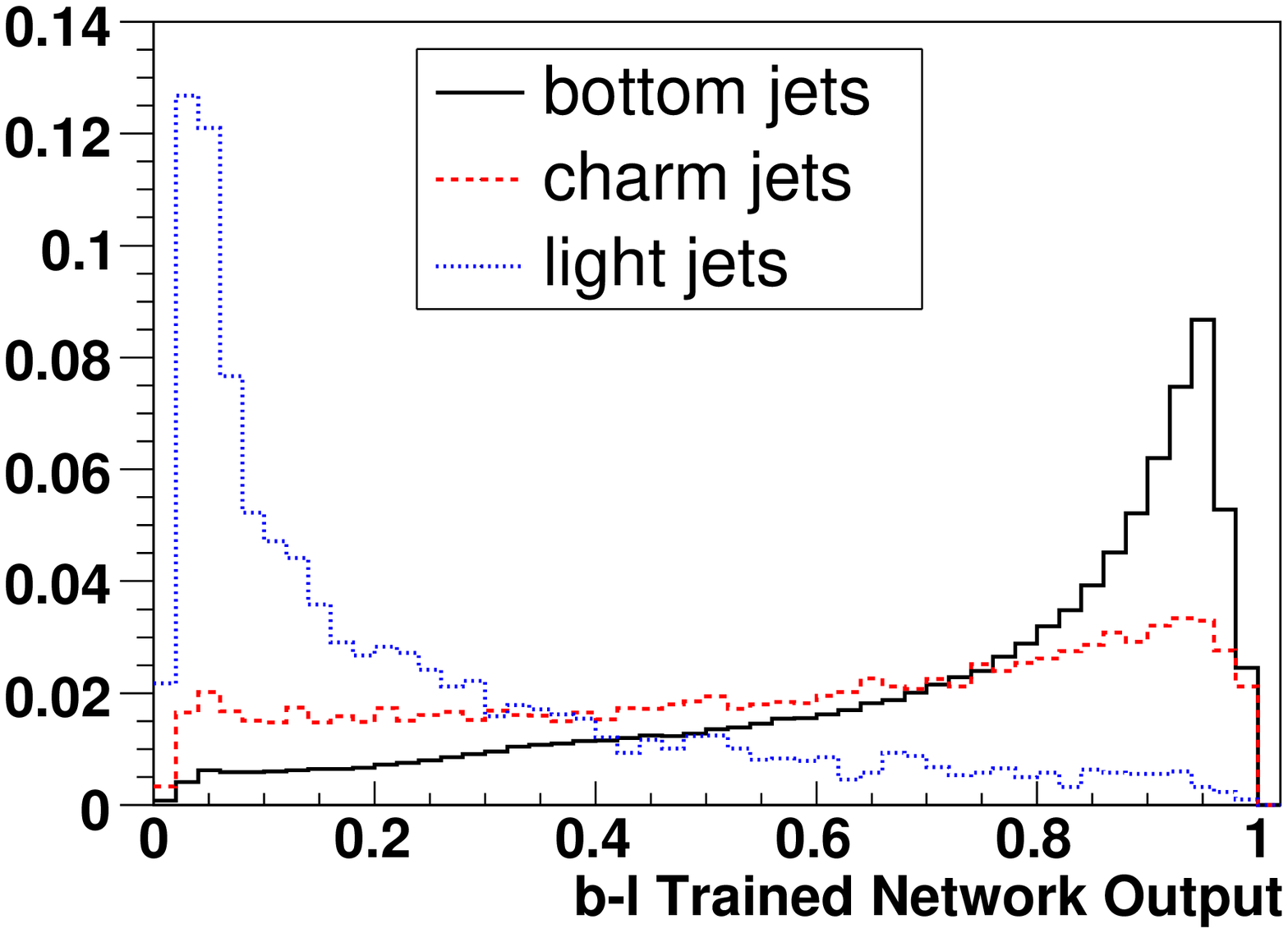}
  \includegraphics[width=0.48\textwidth]{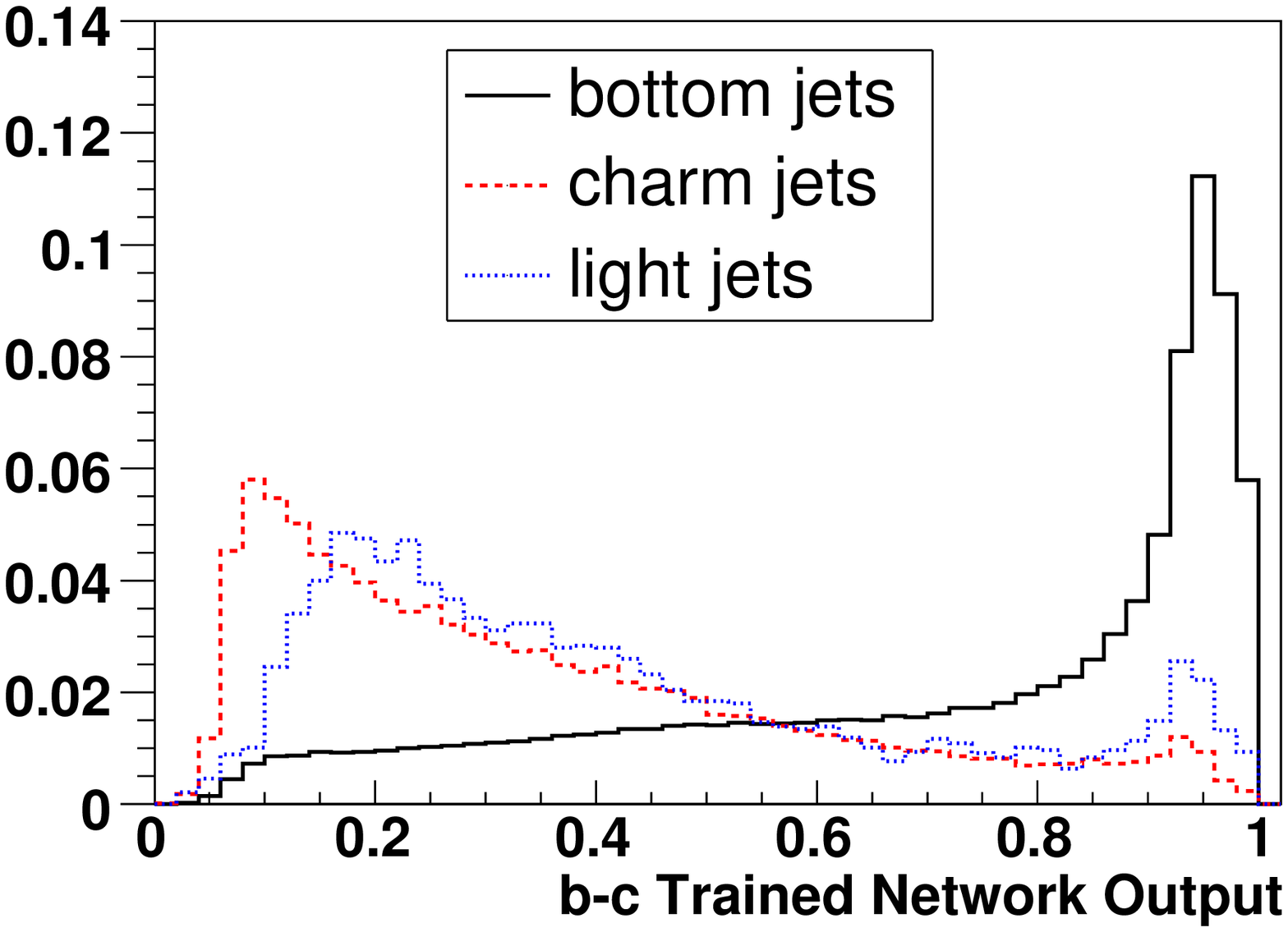}
\caption{Neural network outputs obtained from trainings of $b$ {\it
    vs.} $l$ jets (left) and $b$ {\it vs.} $c$ jets (right). Output
  distributions for $b$, $c$ and $l$ jets are shown in solid, dashed,
  and dotted lines, respectively.
 \label{fig:NNoutput}}
\end{center}
\end{figure}

The neural networks take as input the 16 variables listed in
Table~\ref{tbl:NNvariable}.  These variables are chosen primarily
because the $b$-quark jets have higher track multiplicity, larger
invariant mass, longer lifetime and a harder fragmentation function
than $c$- and $l$-quarks jets. The track parameters and $L_{xy}$
significance are good discriminators for $b$-jets. The vertex
$p_T^{VTX}$ and invariant mass $M_{VTX}$ are useful variables for
identifying $l$-jets; however $c$-jets have $p_T$ spectra similar to
$b$-jets. Pseudo-$c\tau$ ($L_{xy}\times M_{VTX}/p_T^{VTX}$), the
vertex fit $\chi^2$, and the track-based probability of a jet to come
from the primary vertex are the best discriminators. The outputs of
the two neural networks are shown in Fig.~\ref{fig:NNoutput}.
 
\begin{figure}[hbtp]
  \begin{center}
    \includegraphics[width=0.48\textwidth]{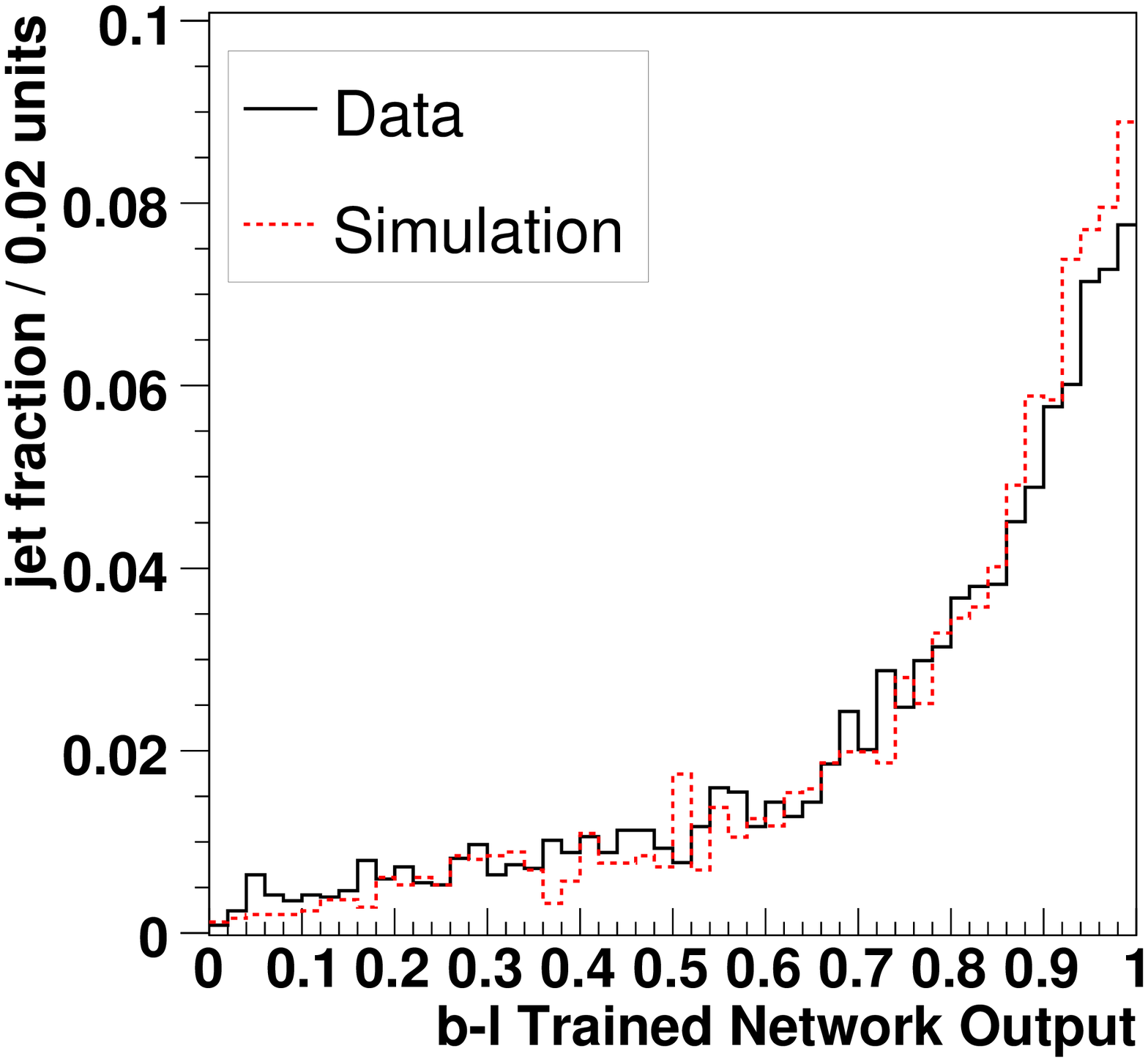}
    \includegraphics[width=0.48\textwidth]{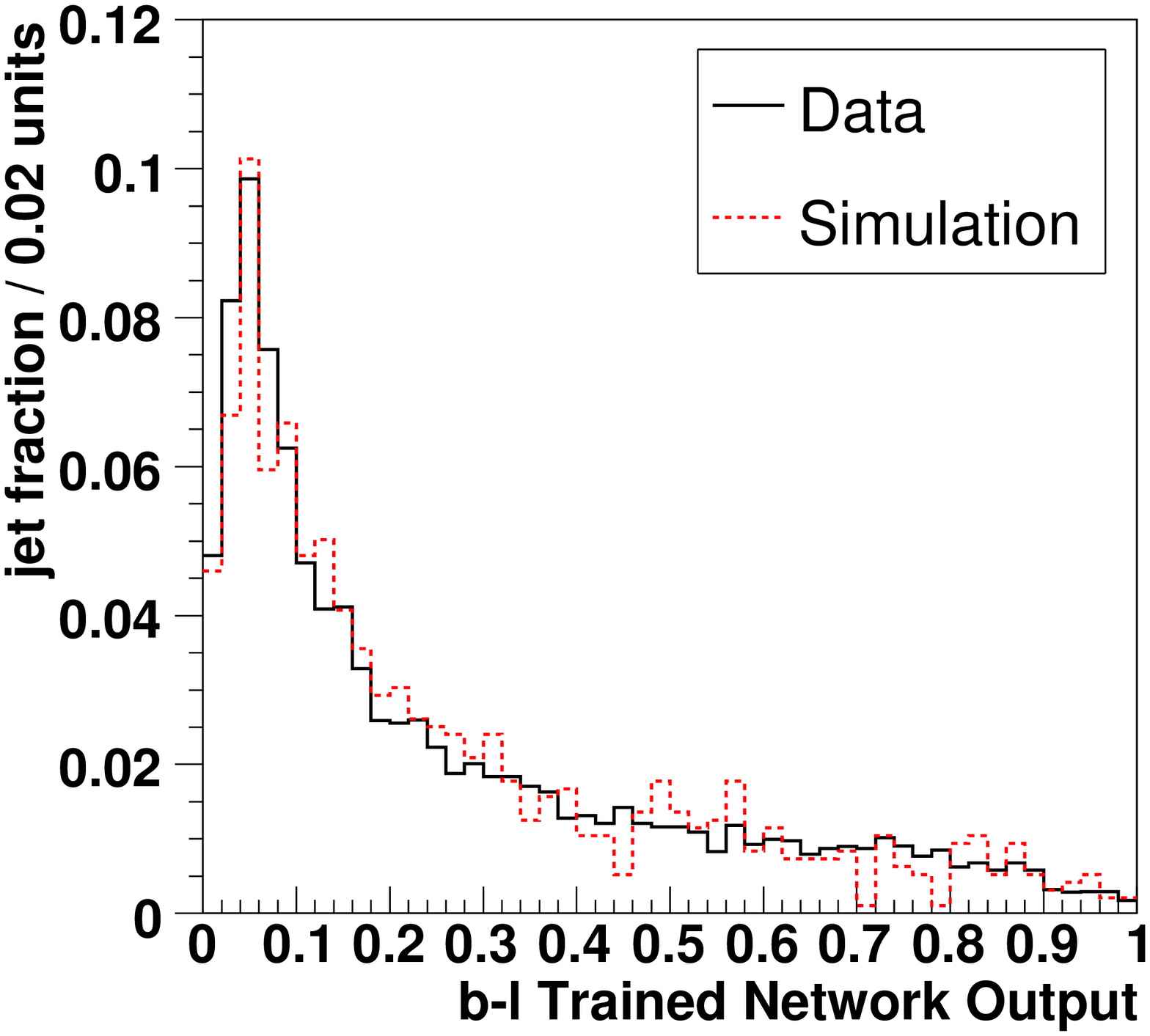}
    \caption{Comparisons of NN $b$-tag output in data~(solid line),
      and Monte Carlo~(dashed line) for {\sc secvtx}-tagged
      heavy-flavor-enriched jets (left) and tagged light-flavor jets
      (right).
    \label{fig:nnbtag}}
  \end{center}
\end{figure}

The NN $b$-tagger is validated by comparing the performance on data
and Monte Carlo events.  The NN output from $b-l$ network on a sample
of {\sc secvtx} tagged heavy-flavor jets from events with an electron
candidate with $E_T>8\,\mathrm{GeV}$ electron data and from the
corresponding Monte Carlo sample are shown in Fig.~\ref{fig:nnbtag},
as are the outputs of the $b-l$ network on tagged light-flavor jets
from data and Monte Carlo\footnote{A small but purified
$b$-jet sample is obtained by requiring a soft lepton in the jet.}.
Figure~\ref{fig:nnbtag} shows the good agreement in NN $b$-tagger
performance between data and Monte Carlo.
 
We tune the cut value for $90\%$ $b$ efficiency (after the {\sc secvtx} efficiency),
corresponding to a value of $\mathrm{NN}_{b-l}=0.182$ and
$\mathrm{NN}_{b-c}=0.242$.  The data-to-Monte-Carlo scale factor,
measured from the electron sample, is $0.97\pm0.02$. Note that this is
an additional scale factor with respect to the {\sc secvtx} efficiency scale
factor because all of the jets under consideration have already been
tagged by {\sc secvtx}.  At these cut values, the NN filter
rejects 65\% of light-flavor jets and about 50\% of the $c$ jets while
keeping 90\% of $b$-jets after being tagged by {\sc secvtx}.
 
\begin{table*}[htbp]
 \begin{center}
 \begin{tabular}{ll}
  \hline \hline
  \multicolumn{1}{c}{{\sc secvtx} variable} & \multicolumn{1}{c}{{\sc secvtx}-independent variable} \\
  \hline
   Number of tracks in fitted vertex &  Number of good
      tracks\\

   Vertex fit $\chi^2$  &  Jet Probability~\cite{Abulencia:2006kv} \\

   Transverse decay length ($L_{xy}$)  &
       Reconstructed mass of pass 1 tracks            \\

   $L_{xy}$ significance ($L_{xy}/\sigma_{L_{xy}}$) &
    Reconstructed mass of pass 2 tracks \\
   Vertex Mass ($M_{\mathrm{vtx}}=\sqrt{(\sum |\bm{p}_{\mathrm{vtx}}|)^2 - (\sum  \bm{p}_{\mathrm{vtx}})^2}$)
  &   Number of pass 1 tracks \\
   Pseudo-$c\tau$ ($L_{xy}\times M_{\mathrm{vtx}}/p_T^{\mathrm{vtx}}$)
  
  &  Number of pass 2 tracks \\
  
   $p_T^{\mathrm{vtx}}/(\sum_{\mathrm{good\  tracks}} p_T)$
  &  $\sum_{\mathrm{Pass1\ track}}p_T/p_T^{jet}$ \\
   Vertex pass number (pass 1 or 2)
  &
       $\sum_{\mathrm{Pass2\ track}}p_T/p_T^{jet}$ \\
  \hline \hline
   \end{tabular}
  \caption{Input variables used in the NN $b$-tagging
  filter.  The variables in the first column are properties of the
  identified secondary vertex, while variables in the second column
  are jet properties independent of any identified vertex.}
  \label{tbl:NNvariable}
  \end{center}
 \end{table*}

\section{Background}\label{sec:bkg}

The final state signature from $WH\rightarrow \ell\nu b\bar b$ production
can also be reached by other production processes.  The main background
processes are $W$+jets production, $t\bar t$ production, and
non-$W$ QCD multijet production.  Several electroweak production
processes also contribute with smaller background rates.
In the following subsections the contribution from each
background source is calculated in detail.

\subsection{Non-$W$ QCD Multijet}
\label{sec:nonW}

Events from QCD multijet production sometimes mimic the
$W$-boson signature with fake leptons or fake \MET .  Non-$W$
leptons are reconstructed when a jet passes the lepton selection
criteria
or a heavy-flavor jet produces leptons via semileptonic decay.
Non-$W$ \MET\ can be observed via mismeasurements of energy or
semileptonic decays of heavy-flavor quarks. It is difficult to model
and produce the former class of events in detector simulation since
the reasons for mismeasurement are not known quantitatively.  Instead,
we estimate the contribution of non-$W$ events directly from the data
sample before $b$-tagging is applied.

Generally, the bulk of non-$W$ events are characterized by a
non-isolated lepton and small \MET .  Lepton isolation $I$ is defined
as the ratio of calorimeter energy inside a cone of $\Delta R = 0.4$
about the lepton to the lepton energy itself.  The quantity $I$ is
small if the lepton is well-isolated from the rest of the event, as
typified by a true leptonic $W$ decay.  This feature is used to
extrapolate the expected non-$W$ contribution into our signal region,
namely, small $I$ and large \MET .  The following 4 sideband sectors
are used for the extrapolation: $I>0.2$ and \MET $<$ 15 GeV (region
A), $I<0.1$ and \MET $<$ 15 GeV (region B), $I>0.2$ and \MET $>$ 20
GeV (region C), and $I<0.1$ and \MET $>$ 20 GeV (region D).
Here, region $D$ corresponds to the signal region.  In extracting the
non-$W$ background contribution from data, we make the following two
assumptions: lepton isolation and \MET\ are uncorrelated in non-$W$
events, and the $b$-tagging rate is not dependent on \MET\ in non-$W$
events.  The level at which these assumptions are justified determines
the assigned uncertainty.

With the first assumption, the number of non-$W$
events~($N_D^{\mathrm{non-}W}$) and their relative fraction in the
signal region before requiring $b$-tagging~($f_{\mathrm{non-}W}$)
obey the following relations:
\begin{eqnarray}
 N_D^{\mathrm{non-}W} &=& \frac{N_B \times N_C}{N_A}, \label{eq:NnonW}\\
 f_{{\rm non}-W} &=&  \frac{N_D^{\mathrm{non-}W}}{N_D} =
\frac{N_B \times N_C}{N_A \times N_D},
  \label{eq:fNonW}
\end{eqnarray}
where $N_i\ (i=A,B,C,D)$ are the number of pretag events in each
sideband region.  The number of pretag events has been corrected for
known sources of prompt leptons.  By invoking the second assumption,
the {\sc secvtx} $b$-tagging efficiency obtained in region $B$ can be
applied to the signal region $D$.  Here we define an event tagging
efficiency per taggable jet (one with at least two good {\sc secvtx}
tracks) as follows:
\begin{equation}
 r_B = \frac{N_B^{\rm (tagged\  event)}}{N_B^{\rm (taggable\  jet)}},
  \label{eq:brate}
\end{equation}
where $N_B^{\rm (tagged\ event)}$ and ${N_B^{\rm (taggable\ jet)}}$
are the number of tagged events and taggable jets in region $B$,
respectively.  Then the number of non-$W$ background in region $D$
after {\sc secvtx} $b$-tagging($N_D^{+\mathrm{non-}W}$) is obtained by using
the ``Tag Rate'' relation:
 \begin{equation}
  N_D^{+\mathrm{non-}W} = f_{{\rm non}-W}\times r_B \times N_D^{\rm (taggable\
   jets)}.
   \label{eq:nonWTagRate}
 \end{equation}
It is also possible to estimate non-$W$ contribution
 solely from the {\sc secvtx}-tagged sample as:
\begin{equation}
 N'^{+\mathrm{non-}W}_D = \frac{N_B^{+} \times N_C^{+}}{N_A^+},
  \label{eq:nonWTagged}
\end{equation}
where $N_X^+(X=A,B,C,D)$ in the ``Tagged Method'' are the number of
events with positive tags.  These methods are data-based techniques,
so the estimates could also contain other background processes.  The
contributions from $t\bar{t}$ and $W$+jets events to each sideband
region are subtracted according to the calculated cross sections for
those processes, including the appropriate tagging efficiencies.

To validate the four-sector method and estimate their systematic
uncertainties, we vary the boundaries of the four regions and divide
the $I$ and \MET\ sidebands into two E ($0.1<I<0.2$ and \MET $>20$
GeV) and F ($I<0.1$ and 15$<$\MET $<$ 20 GeV) sidebands.  The observed
deviations imply a 25\% systematic uncertainty in the non-$W$
background yield, assigned conservatively for both the pretag and
tagged estimates.

The independent estimates from the tag rate method
(Eq.~\ref{eq:nonWTagRate}) and the tagged method
(Eq.~\ref{eq:nonWTagged}) are combined using a weighted average.  The
result from the tagged method gives a slightly higher estimate than
the tag rate method, but the two results are consistent within the
25\% uncertainty.

A non-$W$ rejection factor associated with the NN
$b$-tagging filter is measured from data in region $C$. Region $C$ has
event kinematics similar to non-$W$ events in the signal region
$D$ because lepton isolation is the only difference between the two
regions. The non-$W$ estimate calculated before applying NN
$b$-tagging is scaled by this NN rejection factor; this assumes the NN
filter is uncorrelated with the isolation.

The non-$W$ estimate for events with at least two {\sc secvtx} tags is
obtained by measuring the ratio of the number of events with at least
one $b$-tag to the number with at least two $b$-tags in sideband regions and
applying the ratio to the estimate of tagged non-$W$ events in the
signal region D.

\subsection{Mistagged Jets}
The rate at which {\sc secvtx} falsely tags light-flavor jets is
derived from generic jet samples in varying bins of $\eta$, $\phi$,
jet $E_T$, track multiplicity, and total event $E_T$ scalar sum.  Tag
rate probabilities are summed for all of the taggable jets in the
event, jets with at least two tracks well measured in the silicon
detector.  Since the double-mistag rate is small, this sum is a good
approximation of the single-tag event rate. Negative mistags -- tags
with unphysical negative decay length due to finite tracking
resolution -- are assumed to be a good estimate of falsely tagged
jets, independent to first order of heavy flavor content in the
generic jet sample.  The systematic uncertainty on the rate is largely
due to self-consistency in the parameterization as applied to the
generic jet sample.  The positive mistag rate is enhanced relative to
the negative tag rate by light-flavor secondary vertices and material
interactions in the silicon detectors.  As a result, the positive
mistag rate is corrected by multiplying the negative mistag rate by a
factor of $1.37\pm0.15$.  This factor is measured in a control sample
by fitting the asymmetry in the vertex mass distribution of positive
tags over negative tags~\cite{Abulencia:2006in}.
An additional
correction factor of $1.05\pm0.03$ is applied for data collected after
December 2004, when the Tevatron beam position changed slightly. The
mistag rate per jet is applied to events in the $W$+jets sample.  The
total estimate is corrected for the non-$W$ QCD fraction and also the
top quark contributions to the pretag sample.  To estimate the mistag
contribution in NN-tagged events, we apply the light flavor rejection
power of the NN filter $0.35\pm0.05$ as measured using light-flavor
jets from various data and simulated samples.

\subsection{$W$+Heavy Flavor}

The $Wb\bar{b}$, $Wc\bar{c}$, and $Wc$ states are major background
sources of secondary vertex tags. 
Large theoretical uncertainties exist for the overall
normalization in part because current Monte Carlo
programs generate $W$+heavy-flavor events only to leading order.
Consequently, rates for these processes are normalized to data. 
The contribution from true heavy-flavor production in
$W$+jet events is determined from measurements of the heavy-flavor
event fraction in $W+$jet events and the $b$-tagging efficiency for
those events, as explained below.

The fraction of $W$+jets events produced with heavy-flavor jets has
been studied extensively using an {\sc alpgen}~+~{\sc herwig}
combination of Monte Carlo
programs~\cite{Mangano:2002ea,Corcella:2001wc}.  Calculations of the
heavy-flavor fraction in {\sc alpgen} have been calibrated using a jet
data sample, and measurements indicate a scaling factor of $1.5\pm0.4$
is necessary to make the heavy-flavor production in Monte Carlo match
the production in multijet data~\cite{Acosta:2004hw}. The final
results of heavy-flavor fractions are obtained as shown in
Table~\ref{tbl:HFfrac}.  In the table, 1B and 1C refer to the case in
which only one of the heavy-flavor jets is detected; this happens when
one jet goes out of the detector coverage or when two parton jets
merge into the same reconstructed jet. Similarly, 2B and 2C refer to
the case in which both of the heavy-flavor jets are observed.

For the tagged $W$+heavy flavor background estimate, the heavy-flavor fractions
and tagging rates given in Tables~\ref{tbl:HFfrac}
and~\ref{tbl:HF_tageff} are multiplied by the number of pretag
$W$+jets candidate events in data, after correction for the
contribution of non-$W$ and $t\bar{t}$ events to the pretag sample.

The previous CDF analysis using 319~pb$^{-1}$ of data provided some
evidence that the disagreement between the predicted and observed
numbers of $W+$1 jet and $W+$2 jet events is due to the heavy-flavor
fraction~\cite{Abulencia:2006aj}. In this analysis, an updated
correction factor of $1.2\pm0.2$, obtained by fitting tagged $W$+1 jet
events only, is applied to the heavy-flavor fraction.  The $W+$ heavy
flavor background contribution is obtained by the following
relation:
\begin{equation}
  N_{W+{\rm HF}} = f_{{\rm HF}} \cdot \epsilon _{{\rm tag}} \cdot
  \left [ N_{\rm pretag}\cdot (1-f_{{\rm non}-W}) - N_{{\rm TOP}} - N_{{\rm
  EWK}}\right ],
\end{equation}
where $f_{HF}$ is the heavy-flavor fraction, $\epsilon_{\rm tag}$ is
the tagging efficiency, $N_{\rm TOP}$ is the expected number of $t\bar
t$ and single top events, and $N_{\rm EWK}$ is the expected number of
$WW$, $WZ$, and $Z$ boson events.

\begin{table*}
  \begin{center}
  \begin{tabular}{ccccc}
    \hline \hline
    Jet Multiplicity  & 1 jet & 2 jets & 3 jets & $\geq4$ jets \\
    \hline
    $Wb\bar b$ (1B) (\%)&        1.0 $\pm$     0.3 &        1.4 $\pm$     0.4 &        2.0 $\pm$     0.5 &        2.2 $\pm$     0.6 \\
    $Wb\bar b$ (2B) (\%)&      -   &        1.4 $\pm$     0.4 &        2.0 $\pm$     0.5 &        2.6 $\pm$     0.7 \\
    $Wc\bar c$ (1C) (\%)&        1.6 $\pm$     0.4 &        2.4 $\pm$     0.6 &        3.4 $\pm$     0.9 &        3.6 $\pm$     1.0 \\
    $Wc\bar c$ (2C) (\%)&      -   &        1.8 $\pm$     0.5 &        2.7 $\pm$     0.7 &        3.7 $\pm$     1.0 \\
    $Wc$ (\%)&        4.3 $\pm$     0.9 &        6.0 $\pm$     1.3 &        6.3 $\pm$     1.3 &        6.1 $\pm$     1.3 \\
    \hline \hline
  \end{tabular}
  \caption{The heavy-flavor fractions, given in percent, for the $W$ + jets sample. The results
   from {\sc alpgen} Monte Carlo have been scaled by the data-derived
   calibration factor of $1.5\pm 0.4$. ($Wc$ fractions have not been
   rescaled.) }
   \label{tbl:HFfrac}
  \end{center}
\end{table*}

\begin{table*}
  \begin{center}
\begin{tabular}{ccccc}
\hline\hline
                     Jet Multiplicity &                     1 jet &                     2 jets &                     3 jets &               $\geq4$ jets \\
      \hline
      \multicolumn{5}{c}{$\geq1$ {\sc secvtx} $b$-tag (\%)}\\
      $Wb\bar b$ (1B) &       33.2 $\pm$     2.4 &       34.5 $\pm$     2.5 &       36.7 $\pm$     2.6 &       40.2 $\pm$     2.9 \\
      $Wb\bar b$ (2B) &       -                  &       51.3 $\pm$     3.6 &       54.1 $\pm$     3.8 &       55.1 $\pm$     3.9 \\
      $Wc\bar c$ (1C) &        6.2 $\pm$     0.9 &        8.0 $\pm$     1.1 &        9.7 $\pm$     1.4 &       11.6 $\pm$     1.6 \\
      $Wc\bar c$ (2C) &       -                  &       14.4 $\pm$     2.0 &       17.0 $\pm$     2.4 &       17.8 $\pm$     2.5 \\
      $Wc$      &        8.9 $\pm$     1.3 &        8.7 $\pm$     1.2 &        7.6 $\pm$     1.1 &        3.4 $\pm$     0.5 \\
      \hline

     \multicolumn{5}{c}{$\geq1$ {\sc secvtx} and NN $b$-tag (\%)}\\
 
      $Wb\bar b$ (1B) &       29.9 $\pm$     2.1 &       31.8 $\pm$     2.3 &       34.1 $\pm$     2.4 &       35.9 $\pm$     2.6 \\
      $Wb\bar b$ (2B) &       -                  &       47.2 $\pm$     3.4 &       51.5 $\pm$     3.7 &       51.3 $\pm$     3.6 \\
      $Wc\bar c$ (1C) &        3.8 $\pm$     0.5 &        5.5 $\pm$     0.8 &        6.1 $\pm$     0.9 &        6.4 $\pm$     0.9 \\
      $Wc\bar c$ (2C) &        -                 &        9.9 $\pm$     1.4 &        8.6 $\pm$     1.2 &        9.5 $\pm$     1.4 \\
      $Wc$ &        5.0 $\pm$     0.7 &        4.6 $\pm$     0.7 &        3.1 $\pm$     0.4 &        3.4 $\pm$     0.5 \\
\hline
 \multicolumn{5}{c}{$\geq2$ {\sc secvtx} $b$-tag (\%)}\\
      $Wb\bar b$ (2B) &         -                 &        9.7 $\pm$     0.7 &       13.6 $\pm$     1.0 &       11.5 $\pm$     0.8 \\
      $Wc\bar c$ (2C) &        -                  &        1.2 $\pm$     0.2 &        0.8 $\pm$     0.1 &        0.9 $\pm$     0.1 \\
 \hline\hline
\end{tabular}
\caption{The $b$-tagging efficiencies in percent for various $b$-tagging strategies
   on individual $W$+heavy-flavor processes.  Categories $1B$, $2B$
   refer to number of taggable $b$-jets in the events, with similar
   categories for charm jets.  Those numbers include the effect of the
   data-to-Monte Carlo scale factors 
   algorithm and the neural network filter.}
\label{tbl:HF_tageff}
  \end{center}
\end{table*}

\subsection{Top and Electroweak Backgrounds}
Production of both single top quark and top-quark pairs contribute to
the tagged lepton+jets sample.  Several electroweak boson production
processes also contribute.  $WW$ pairs can decay to a lepton, neutrino
as missing energy, and two jets, one of which may be charm. $WZ$
events can decay to the signal $Wb\bar b$ or $Wc\bar c$ final
state. Finally, $Z\rightarrow\tau^+\tau^-$ events can have one
leptonic $\tau$ decay and one hadronic decay. The leptonic $\tau$
decay gives rise to a lepton + missing transverse energy, while the
hadronic decay yields a narrow jet of hadrons with a non-zero
lifetime.

The normalization of the diboson and single top backgrounds are based
on the theoretical cross sections listed in Table~\ref{tbl:xsec}, the
luminosity, and the acceptance and $b$-tagging efficiency derived from
Monte Carlo
events~\cite{Campbell:2002tg,Acosta:2004uq,Cacciari:2003fi,Harris:2002md}. The
acceptance is corrected for lepton identification, trigger
efficiencies, and the $z$ vertex cut. The tagging efficiency is always
corrected by the $b$-tagging scale factor.

\begin{table}
  \begin{center}
    \begin{tabular}{cc}
      \hline
      Theoretical Cross
    Sections\\ \hline $WW$ & 12.40 $\pm$ 0.80 pb \\ $WZ$ & 3.96 $\pm$
    0.06 pb \\ $ZZ$ & 1.58 $\pm$ 0.02 pb \\ Single top $s$-channel &
    0.88 $\pm$ 0.05 pb\\ Single top $t$-channel & 1.98 $\pm$ 0.08 pb\\
    $Z\rightarrow \tau^+\tau^- $ & 320 $\pm$ 9 pb\\ $t\bar{t}$ & 6.7
    $^{+0.7} _{-0.9}$ pb \\
    \hline\hline
    \end{tabular}
    \caption{Theoretical
    cross sections and uncertainties for the electroweak and single top
    backgrounds, along with the theoretical cross section for
    $t\bar{t}$ at $m_t = 175\,\mathrm{GeV}/c^2$. The cross section of $Z^{0}
    \rightarrow \tau^+\tau^-$ is obtained in the dilepton mass range
    $m_{\tau\tau}>30\,\mathrm{GeV}/c^2$ together with a $k$-factor (NLO/LO) of
    1.4.
    \label{tbl:xsec}}
  \end{center}
\end{table}

\subsection{Summary of Background Estimate}
\begin{figure}[htbp]
  \begin{center}
    \includegraphics*[width=0.48\textwidth]{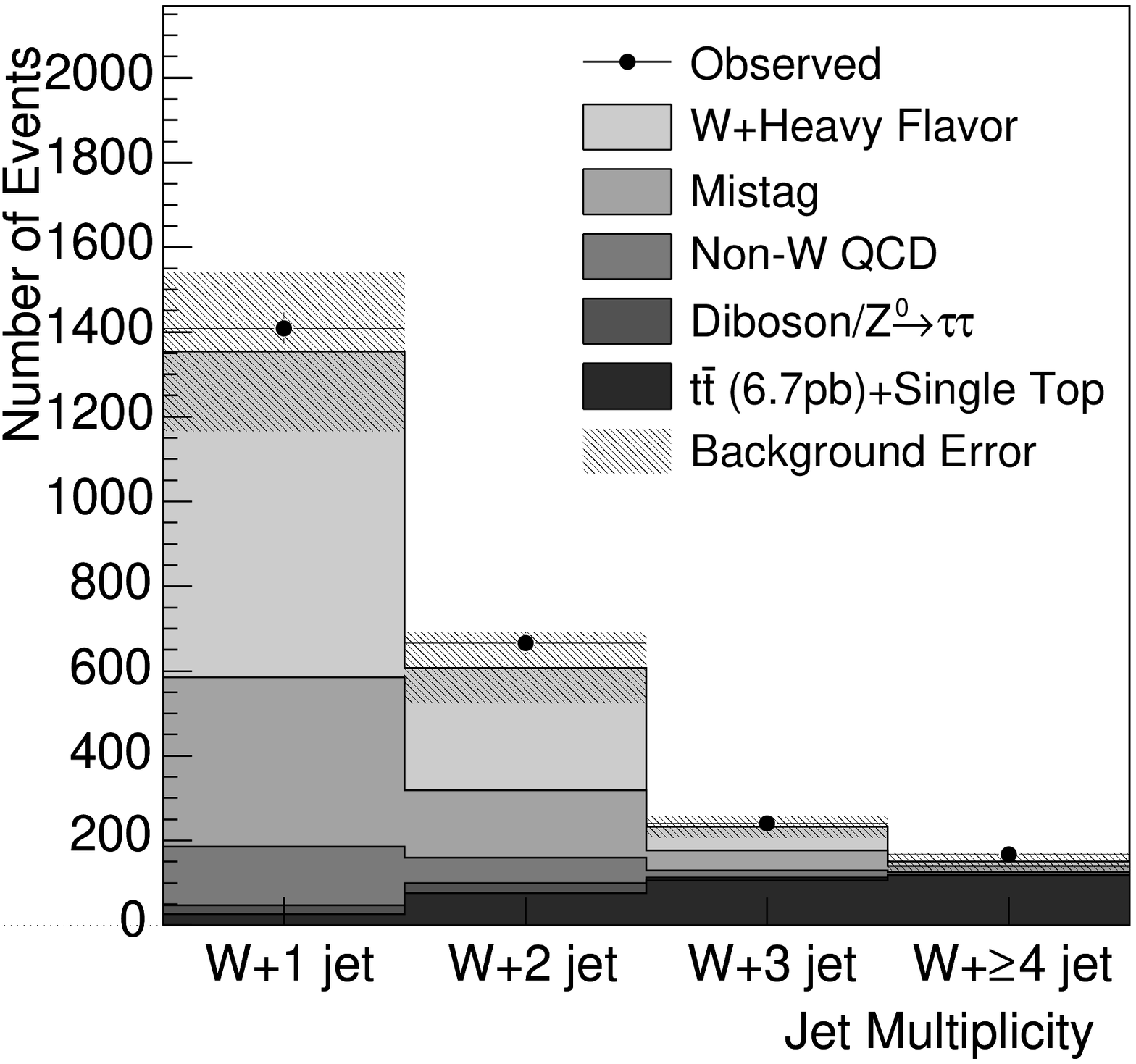}
    \includegraphics*[width=0.48\textwidth]{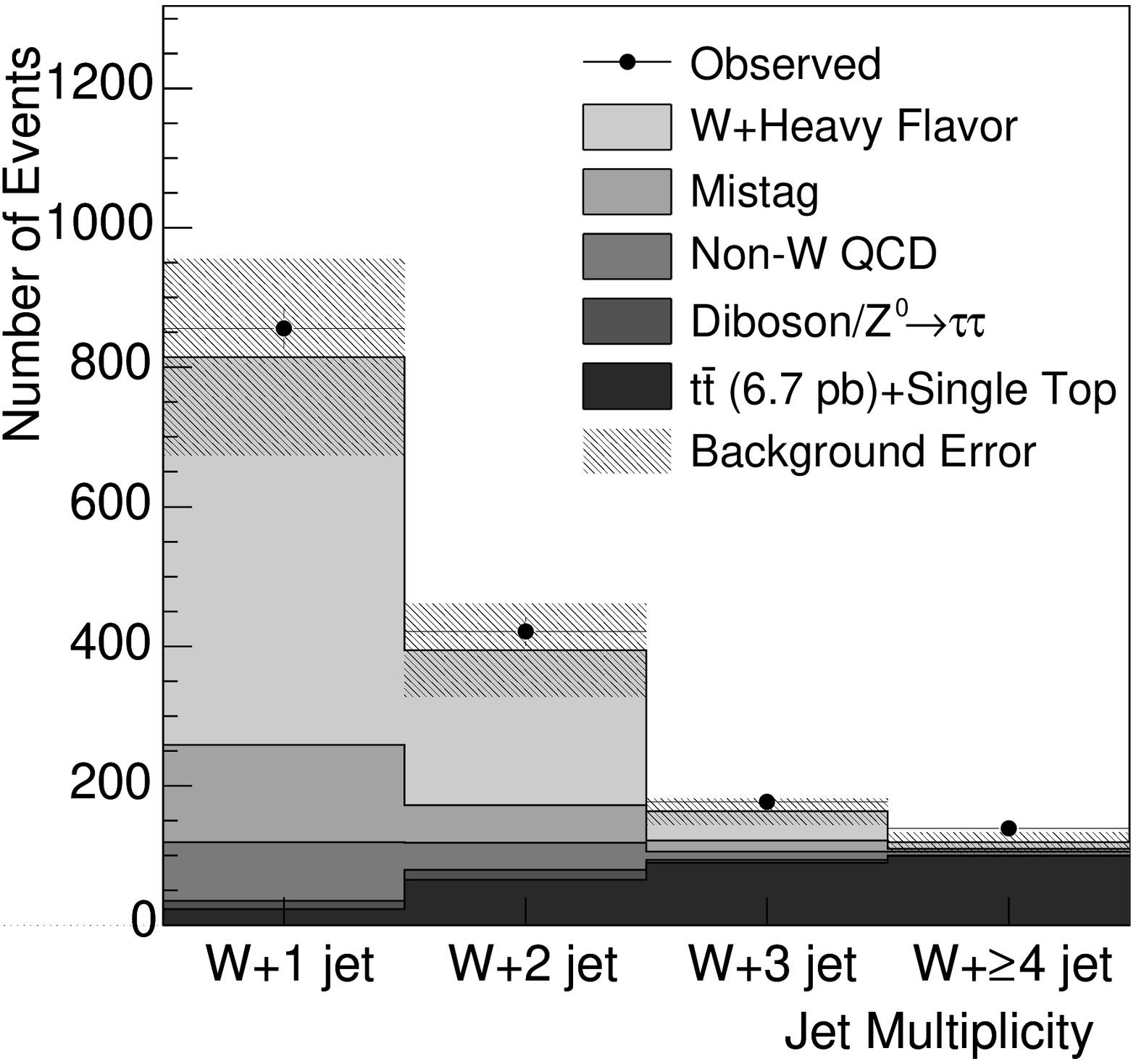}
    \caption{Number of events as a function of jet multiplicity
       for events with exactly one {\sc secvtx} $b$-tag before(left) and after(right)
       applying the NN $b$-tagging requirement.
    \label{fig:Njets_singletag_0d0h0i_All}}
  \end{center}
\end{figure}

\begin{figure}[htbp]
  \begin{center}
    \includegraphics*[width=0.48\textwidth]{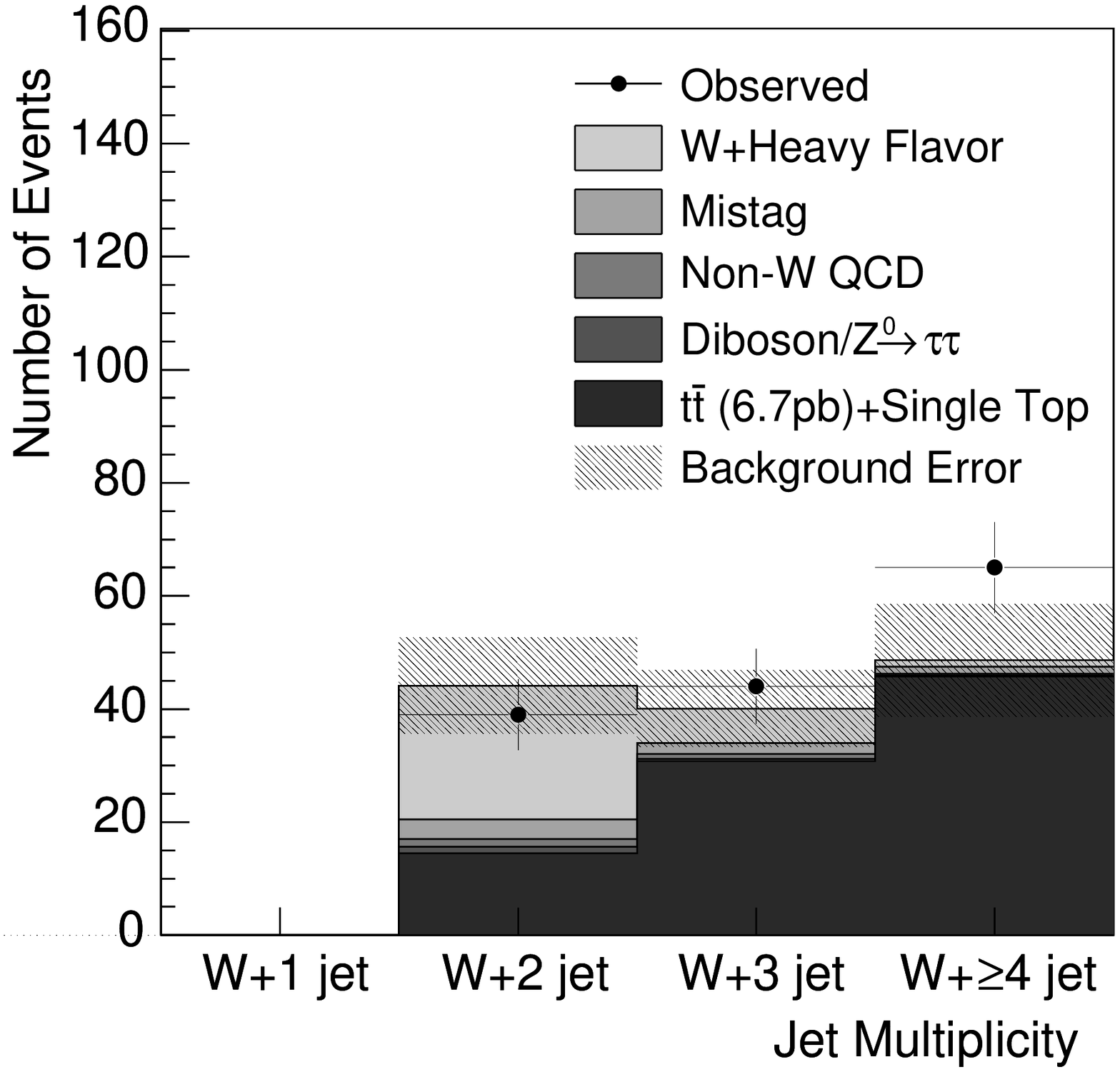}
    \caption{Number of events as a function of jet multiplicity
       for events with at least two {\sc secvtx} $b$-tagged jets.
    \label{fig:Njets_secvtx_0d0h0i_All}}
  \end{center}
\end{figure}

We have described the contributions of individual background sources
to the final background estimate.  The background estimates for the
condition of exactly one b-tagged jet after applying the NN
filter and at least two {\sc secvtx} $b$-tagged jets are summarized in
Tables~\ref{tbl:Njets_snntag_0d0h0i_All}
and~\ref{tbl:Njets_secvtx_0d0h0i_All}.  The estimates are plotted
in Figs.~\ref{fig:Njets_singletag_0d0h0i_All}
and~\ref{fig:Njets_secvtx_0d0h0i_All} for the case of exactly one
$b$-tag before and after applying the NN $b$-tag filter.  The
observed number of events in the data and the SM
background expectations are consistent both before and after NN
$b$-tagging is applied. The same is true for the number of
events with at least two $b$-tagged jets.  (See
Table~\ref{tbl:Njets_secvtx_0d0h0i_All} and
Fig.~\ref{fig:Njets_secvtx_0d0h0i_All}.)

  \begin{table*}
    \begin{center}
\begin{tabular}{ccccc}
\hline \hline
                   Jet Multiplicity &                   1 jet &                   2 jets &                   3 jets &             $\geq4$ jets \\
\hline
Pretag Events &                  94051 &                  14604 &                   2362 &                    646 \\
\hline
                             Mistag &     139.7 $\pm$ 27.3 &      53.9  $\pm$ 10.7 &      15.7 $\pm$  3.1 &       4.2 $\pm$  0.8 \\
                        $Wb\bar{b}$ &     306.9 $\pm$ 106.9 &     144.7 $\pm$ 49.4 &      29.9 $\pm$  9.7 &       6.4 $\pm$  2.5 \\
                        $Wc\bar{c}$ &      63.1 $\pm$ 22.0 &      43.0  $\pm$ 14.7 &       8.7 $\pm$  2.8 &       1.9 $\pm$  0.8 \\
                               $Wc$ &     185.7 $\pm$ 47.2 &      34.4  $\pm$  9.0 &       3.4 $\pm$  0.9 &       0.6 $\pm$  0.2 \\
                  $t\bar{t}$(6.7pb) &       6.9 $\pm$  1.2 &      42.0  $\pm$  6.6 &      84.9 $\pm$ 12.8 &      98.6 $\pm$ 14.3 \\
                         Single Top &      16.7 $\pm$  1.8 &      23.5  $\pm$  2.4 &       4.8 $\pm$  0.5 &       0.8 $\pm$  0.1 \\
Diboson/$Z^{0}\rightarrow \tau^+\tau^-$ &      11.7 $\pm$  2.2 &      14.2  $\pm$  2.3 &       3.9 $\pm$  0.9 &       1.0 $\pm$  0.3 \\
                        non-$W$ QCD &      84.2 $\pm$ 14.1 &      38.9  $\pm$  6.7 &      12.1 $\pm$  2.3 &       5.5 $\pm$  1.2 \\
\hline \hline
                   Total Background &     814.9 $\pm$ 140.7 &     394.4 $\pm$ 66.6 &     163.4 $\pm$ 18.7 &     118.9 $\pm$ 14.9 \\
\hline
        Observed Events &                    856 &                    421 &                    177 &                    139 \\
\hline\hline
\end{tabular}
      \caption{Background
  estimate for events with exactly one {\sc secvtx} $b$-tag that passes the
      NN filter as a function of jet multiplicity.
      \label{tbl:Njets_snntag_0d0h0i_All}}
    \end{center}
  \end{table*}

\begin{table*}
  \begin{center}
\begin{tabular}{ccccc}
\hline \hline
                   Jet Multiplicity &                    2 jets &                   3 jets &             $\geq4$ jets \\
\hline
Observed Events(pretag)             &                   14604 &                   2362 &                    646 \\
\hline
                             Mistag &        3.5 $\pm$  0.5 &       2.0  $\pm$  0.3 &       1.2 $\pm$  0.2 \\
                        $Wb\bar{b}$ &       20.3 $\pm$  7.0 &       5.7  $\pm$  1.8 &       1.0 $\pm$  0.4 \\
                        $Wc\bar{c}$ &        3.3 $\pm$  1.1 &       0.4  $\pm$  0.1 &       0.1 $\pm$  0.04 \\
                               $Wc$ &              -         &             -        &             -         \\
                  $t\bar{t}$ (6.7pb) &       10.4 $\pm$  2.3 &      29.5 $\pm$  6.4 &      45.5 $\pm$  9.9 \\
                         Single Top &        4.2 $\pm$  0.7 &       1.4 $\pm$  0.2 &       0.3 $\pm$  0.1 \\
Diboson/$Z^{0}\rightarrow \tau^+\tau^-$ &        1.2 $\pm$  0.3 &       0.3 $\pm$  0.1 &       0.1 $\pm$  0.1 \\
                        non-$W$ QCD &        1.4 $\pm$  0.3 &       0.9 $\pm$  0.2 &       0.3 $\pm$  0.1 \\
\hline \hline
                   Total Background &       44.2 $\pm$  8.5 &      40.1 $\pm$  6.8 &      48.6 $\pm$  10.0 \\
\hline
        Observed Events &                      39 &                     44 &                     65 \\
\hline\hline
\end{tabular}
    \caption{Background
  estimate for events with at least two {\sc secvtx} $b$-tagged jets as a function of jet multiplicity.
    \label{tbl:Njets_secvtx_0d0h0i_All}}
  \end{center}
\end{table*}

\section{Higgs Boson Signal Acceptance}
\label{sec:Acceptance}
\begin{figure}
\begin{center}
 \includegraphics[width=0.48\textwidth]{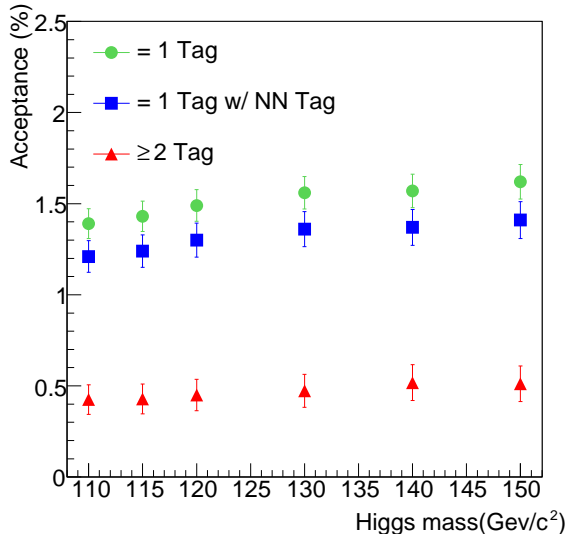}
\end{center}
\caption{The summary of acceptance of the process $WH \rightarrow \ell
 \nu b\bar{b}$ in W+2jet bin for various $b$-tagging strategies
 as a function of Higgs boson mass.
\label{fig:acc2woPHX}}
\end{figure}

The kinematics of the SM $WH\rightarrow \ell\nu b\bar b$ process are
well defined, and events can be simulated accurately by Monte Carlo
programs. The {\sc pythia} program was used to generate the signal
samples~\cite{Sjostrand:2000wi}.
Only Higgs boson masses between 110 and $150\,\mathrm{GeV}/c^2$ are
considered because this is the mass region for which the decay
$H\rightarrow b\bar b$ dominates.  The number of expected
$WH\rightarrow \ell\nu b\bar b$ events $N$ is given by:
\begin{equation}
N = \epsilon \cdot \int {\cal{L}} dt \cdot \sigma (p \bar{p}
 \rightarrow WH)\cdot {\cal B}(H \rightarrow b \bar{b}), \label{eq:ExpEvt}
 \label{eq:WHexpect}
\end{equation}
where $\epsilon$, $\int {\cal{L}}
dt$, $\sigma(p \bar{p} \rightarrow WH)$, and ${\cal B}(H\rightarrow
b\bar{b})$ are the event detection efficiency, integrated luminosity,
production cross section, and branching ratio, respectively.  The
production cross section and branching ratio are calculated to NLO
precision~\cite{Djouadi:1997yw}.  The acceptance
$\epsilon$ is broken down into the
following factors:
\begin{equation}
\epsilon = 
 \sum_{\ell=e,\mu,\tau } \left( \epsilon_{z_0} \cdot
 \epsilon_{\mathrm{trigger}} \cdot \epsilon_{\mathrm{lepton\ ID}}
\cdot \epsilon _{b\mathrm{tag}} \cdot
 \epsilon_{\mathrm{kinematics}} \cdot  {\cal B}(W \rightarrow \ell \nu) \right), \label{eq:sigEff}
\end{equation}
where $\epsilon_{z_0}$, $\epsilon_{\mathrm{trigger}}$,
$\epsilon_\mathrm{lepton\ ID}$, $\epsilon _{b\mathrm{tag}}$, and
$\epsilon _{\mathrm{kinematics}}$ are efficiencies to meet the
requirements of primary vertex, trigger, lepton identification,
$b$-tagging, and kinematics.  The major sources of inefficiency are
the lepton identification, jet kinematics, and $b$-tagging factors;
each is a factor between 0.3 and 0.45.  The factor $\epsilon_{z_0}$ is
obtained from data, and the others are calculated using Monte Carlo
samples.  The total signal acceptances for various $b$-tagging options
including all systematic uncertainties as a function of Higgs boson
mass are shown in Fig.~\ref{fig:acc2woPHX}.

The expected number of 
signal events
is estimated by Eq.~\ref{eq:WHexpect} at each Higgs boson mass point.  The
expectations for various $b$-tagging strategies are shown in
Table~\ref{tbl:ExEvt2woPHX}.  The NN $b$-tagging filter keeps
about 90\% of signal while it removes 35\% of the total
background in $W$+2 jet events as shown in
Fig.~\ref{fig:Njets_singletag_0d0h0i_All}.

\begin{table*}
\begin{center}
    \begin{tabular}{ccccc} \hline\hline
      Higgs Mass & \multicolumn{4}{c}{Expected Signal Events} \\
      (GeV/$c^2$)     & Pretag         & 1 tag            & 1 tag with NNtag   &   $\geq$ 2 tag   \\ \hline
      110        & 4.81$\pm$0.34  & 2.15  $\pm$ 0.18  & 1.87  $\pm$ 0.18  &  0.66 $\pm$ 0.13 \\
      115        & 3.99$\pm$0.28  & 1.80  $\pm$ 0.15  & 1.56  $\pm$ 0.15  &  0.54 $\pm$ 0.11 \\
      120        & 3.23$\pm$0.23  & 1.45  $\pm$ 0.12  & 1.26  $\pm$ 0.12  &  0.44 $\pm$ 0.09 \\
      130        & 2.05$\pm$0.15  & 0.93  $\pm$ 0.08  & 0.81  $\pm$ 0.08  &  0.28 $\pm$ 0.06 \\
      140        & 1.03$\pm$0.07  & 0.46  $\pm$ 0.04  & 0.40  $\pm$ 0.04  &  0.15 $\pm$ 0.03 \\
      150        & 0.40$\pm$0.03  & 0.18  $\pm$ 0.02  & 0.16  $\pm$ 0.02  &  0.06 $\pm$ 0.01 \\
      \hline
    \end{tabular}
\end{center}
\caption{Expected number of $WH\rightarrow \ell \nu b\bar{b}$ signal
  events with systematic uncertainties for various $b$-tagging
  options, where ``tag'' and ``NNtag'' stand for {\sc secvtx}
  $b$-tagging and NN $b$-tagging,
  respectively.}\label{tbl:ExEvt2woPHX}
\end{table*}

The total systematic uncertainty on the acceptance stems from the jet
energy scale, initial and final state radiation, lepton
identification, trigger efficiencies, and $b$-tagging.
A 2\% uncertainty on the lepton identification efficiency is assigned
for each lepton type (CEM electron, CMUP and CMX muon), based on
studies of $Z$ boson events.  For each of the high $p_T$ lepton
triggers, a 1\% uncertainty is measured from backup trigger paths or
$Z$ boson events.  The initial and final state radiation systematic
uncertainties are estimated by changing the parameters related to ISR
and FSR from nominal values to half or double the
nominal~\cite{Abulencia:2005aj}.  The difference from the nominal
acceptance is taken as the systematic uncertainty.  The uncertainty in
the incoming parton energies relies on the eigenvalue uncertainties provided in
the PDF fits.  An NLO version of the PDFs, CTEQ6M, provides a 90\%
confidence interval of each eigenvector~\cite{Pumplin:2002vw}. The
nominal PDF value is reweighted to the 90\% confidence level value,
and the corresponding reweighted acceptance is computed.  The
differences between nominal and reweighted acceptances are added in
quadrature, and the total is assigned as the systematic
uncertainty~\cite{Acosta:2004hw}.

The uncertainty due to the jet energy scale uncertainty
(JES)~\cite{Bhatti:2005ai} is calculated by shifting jet energies in $WH$
Monte Carlo samples by $\pm 1\sigma$.  The deviation from the nominal
acceptance is taken as the systematic uncertainty.  The systematic
uncertainty on the {\sc secvtx} $b$-tagging efficiency is based on the scale
factor uncertainty discussed in Sec.~\ref{sec:btagging}. When NN
$b$-tagging is applied, the scale factor uncertainty
is added to that of {\sc secvtx} in quadrature.  The total systematic
uncertainties for various $b$-tagging options are summarized in
Table~\ref{tbl:Sys2jet}.
    
\begin{table}
\begin{center}
\begin{tabular}{cccc} \hline\hline
source    & \multicolumn{3}{c}{uncertainty (\%)} \\ \hline
          &  1 Tag  &  1 Tag \& NNtag & $\geq$ 2 Tag \\ \hline
Lepton ID & 2.0\%     &  2.0\%           & 2.0\%   \\
Trigger   & $<$1\%    &  $<$1\%          & $<$1\%      \\
ISR       & 1.5\%     &  1.8\%           & 4.3\%       \\
FSR       & 2.8\%     &  3.2\%           & 8.6\%       \\
PDF       & 1.6\%     &  1.7\%           & 2.0\%       \\
JES       & 2.3\%     &  2.3\%           & 3.0\%       \\
$b$-tagging & 3.8\%     &  5.3\%           & 16\%        \\ \hline
Total     & 5.8\%     &  7.2\%           & 19\%      \\  \hline
\end{tabular}
\end{center}
\caption{Systematic uncertainties for various $b$-tagging requirements.
  The labels ``Tag'' and ``NNtag'' refer to {\sc secvtx} and NN $b$-tagging,
  respectively.}
\label{tbl:Sys2jet}
\end{table}

\section{Optimization of Search Strategies}
\label{sec:interpretation}
\begin{figure}[tbp]
  \begin{center}
   \includegraphics[width=0.48\textwidth]{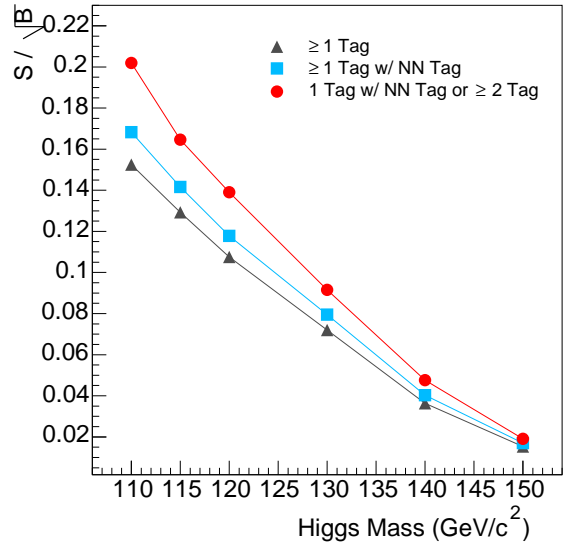}
   \caption{Comparison of significance obtained from various
   $b$-tagging strategies. ``Tag'' and ``NN Tag'' represent {\sc
   secvtx} and NN $b$-tagging, respectively. The filled circles
   correspond to the combined analysis which treats disjoint samples with
   exactly one NN $b$-tag and two {\sc secvtx} tags separately.
   \label{fig:sensitivity}} \end{center}
\end{figure}

The search strategy is optimized by calculating a signal significance
defined as $S/\sqrt{B}$, where $S$ and $B$ are the number of expected
signal and background events. In this analysis, $S$ and $B$ are
counted within a window which gives the best significance in dijet
mass distribution for the particular Higgs mass hypothesis being
considered.  The window itself is optimized by varying the window peak
and width for each $b$-tagging strategy.  A comparison of the
significance for various $b$-tagging options, shown in
Fig.~\ref{fig:sensitivity}, provides an {\it a priori} metric that
predicts which selection gives the best result.

Requiring the NN filter improves the sensitivity by about 10\% in the
sample of events with exactly one $b$ tag.  The significance in
double-tagged events is almost the same as that in events with at
least one tag and no NN filter.  Combining the two results therefore
yields another sensitivity improvement.
This combined use of two separate $b$-tagged samples provides a
significant improvement as shown in Fig.~\ref{fig:sensitivity}.  The total
significance increases by 20\% moving from ``$\geq1$ tag'' to separate categories 
``1 tag w/ NNTag'' and ``$\geq 2$ Tag.''  Therefore, we quote final
results from events having exactly one {\sc secvtx} $b$-tagged jet passing
the neural network filter or at least two {\sc secvtx} $b$-tagged jets.

\section{Limit on Higgs Boson Production Rate}
\label{sec:limit}
\begin{figure}
  \begin{center}
   \includegraphics[width=0.48\textwidth]{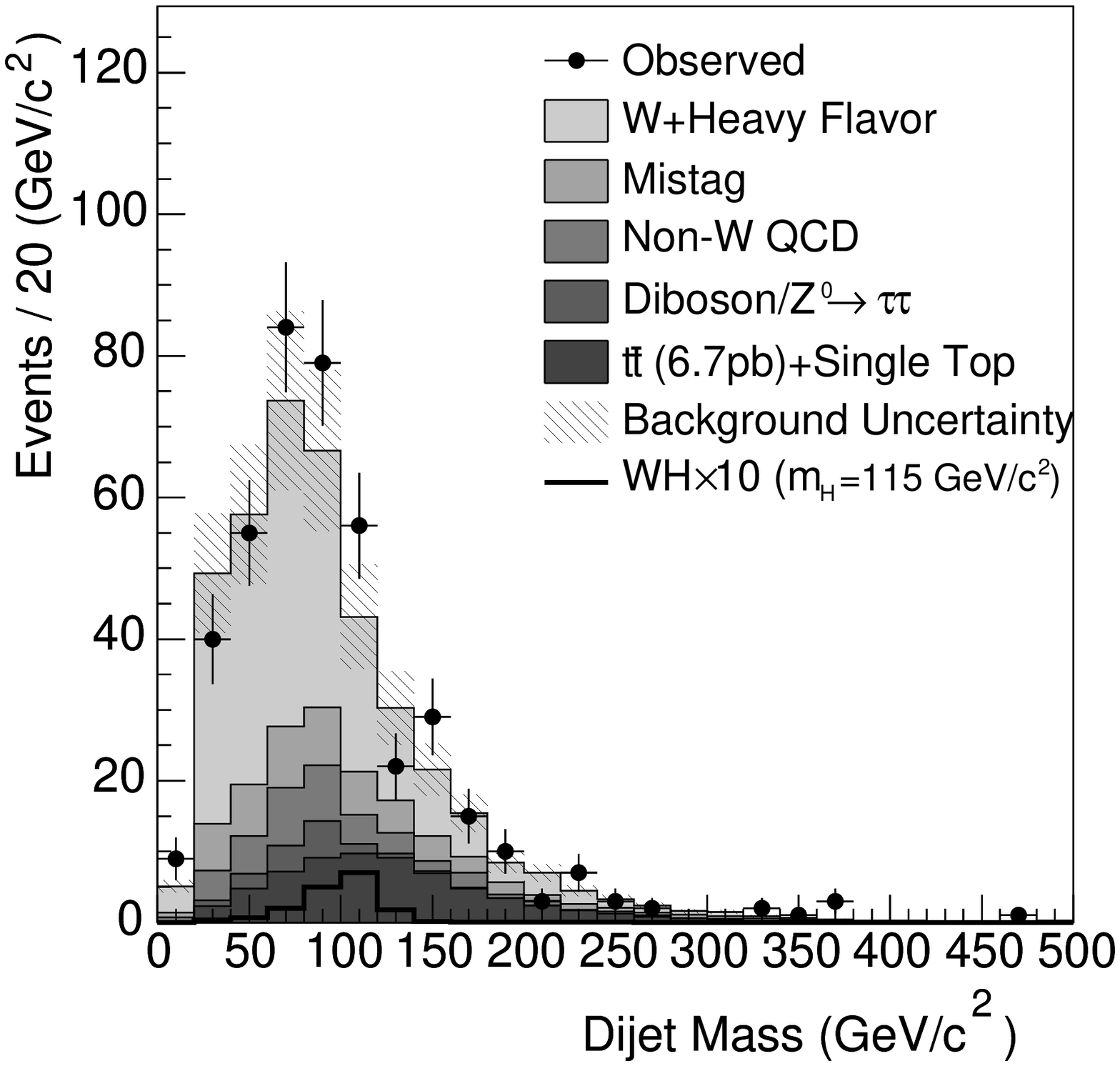}
   \caption{Dijet mass distribution in $W$+2 jets events including
     exactly one {\sc secvtx} $b$-tagged jet that passes the NN
     $b$-tagging filter.  The contributions of the various background sources
     are shown in histogram, while the hatched box on the background
     histogram represents the background uncertainty.
     \label{fig:m2j_snntag}}
  \end{center}
\end{figure}

\begin{figure}
  \begin{center}
    \includegraphics[width=0.48\textwidth]{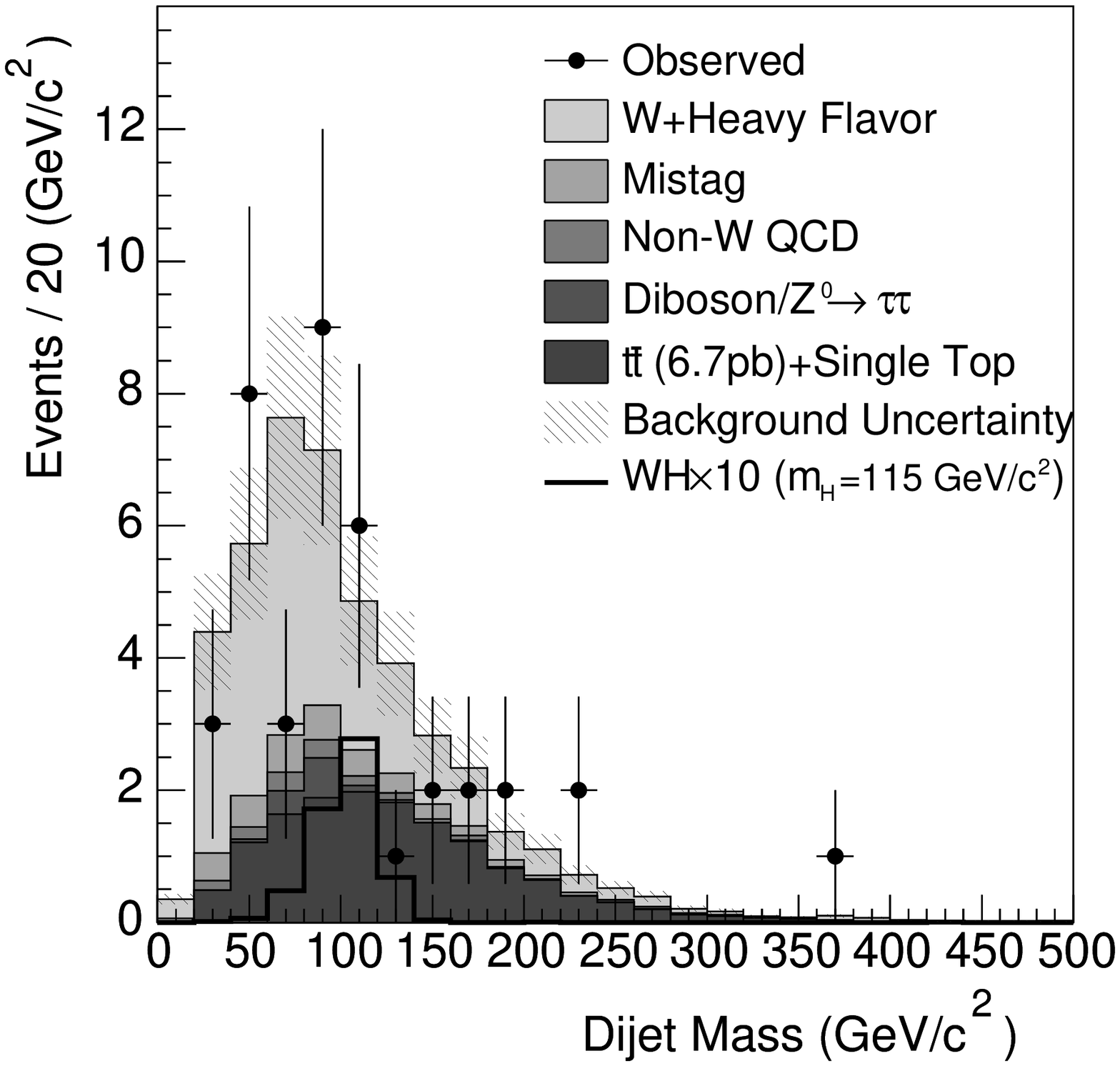}
    \caption{Dijet mass distribution in $W$+2 jets events including at
      least two {\sc secvtx} $b$-tagged jets.
    \label{fig:m2j_secvtx}}
  \end{center}
\end{figure}

As shown in section~\ref{sec:interpretation}, there is no
significant excess number of events over the SM background
expectation.  Because the dijet mass resonance is a useful
discriminant for the Higgs boson signature,
we use a binned likelihood technique to fit the observed dijet mass distributions in Figs.~\ref{fig:m2j_snntag} and~\ref{fig:m2j_secvtx}, and set
an upper limit on the $WH$ production cross section times $H\rightarrow b\bar
b$ branching ratio.


\subsection{Binned Likelihood Technique}
The number of events in each bin follows the Poisson distribution:
\begin{equation}
  P_i (n_i, \mu_i) = \frac{\mu_i ^{n_i} e^{-\mu_i}}{n_i !}
  \label{eq:Poisson} \quad (i=1,2,\cdots, N_{{\rm bin}})
\end{equation}
where $n_i,\mu_i$, and $N_{\mathrm{bin}}$ represent the number of
observed events in the $i$-th bin, the expectation in the $i$-th bin, and the
total number of bins.  The Higgs production hypothesis is constructed
by setting $\mu_i$ to $\mu_i =	s_i + b_i$,
where $s_i$ and $b_i$ are the number of signal and expected background
events in the $i$-th bin.
This quantity $s_i$ can also be written as a product:
\begin{equation}
s_i =  \sigma(p\bar{p}\rightarrow W^{\pm}H)\cdot{\cal B}(H\rightarrow b\bar{b}) \cdot  \epsilon _{WH}\cdot \int {\mathcal{L}}dt \cdot f_i^{WH}
\end{equation}
where $f_i^{WH}$ is the fraction of the
total signal which lies in the $i$-th bin.  
In this case,
$\sigma(p\bar{p}\rightarrow W^{\pm}H)\cdot{\cal B}(H\rightarrow
b\bar{b})$ is the variable to be extracted from data.  An upper limit
on the Higgs boson production cross section times branching ratio
$\sigma(p\bar{p}\rightarrow W^{\pm}H)\cdot {\cal B}(H\rightarrow
b\bar{b})$ is extracted by using a Bayesian procedure with a
likelihood defined by:
\begin{equation}
  L   = \prod _{i=1}^{N_{\mathrm{bin}}} P_i(n_i,\mu_i) = \prod _{i=1}^{N_{\mathrm{bin}}}
    \frac{\mu_i^{n_i}e^{-\mu_i}}{n_i !}.
    \label{eq:likelihood}
\end{equation}
The background prediction $b_i$ includes contributions from the
various background sources described in Sec.~\ref{sec:bkg}:
\begin{equation}
b_i = N^{TOP}f_i^{TOP} + N^{QCD}f_i^{QCD},
\end{equation}
where $f_i^{TOP}$ and $f_i^{QCD}$ are the fractions of the total number of top (including $t\bar t$ and single top)
and QCD backgrounds (including W+jets, non-$W$, and diboson) in mass bin $i$.
There are systematic
uncertainties in the estimates of both the number of signal events and
the expected background.  Such uncertainties modify the likelihood to
be
\begin{eqnarray}
\!\!\!\!\!\!\! L(\sigma \cdot {\cal B}) &=&\int _{N^{QCD}} \int _{N^{TOP}} \int _{N^{WH}}
\prod  _{i=1}^{N_\mathrm{bin}} \frac{\mu_i ^{n_i} e^{-\mu_i}}{n_i!}\nonumber\\
&\times& G(N^{QCD},\sigma^{QCD})  G(N^{TOP},\sigma^{TOP})  G(N^{WH},\sigma^{WH})
dN^{QCD} dN^{TOP}dN^{WH}
\label{eq:comboLike}
\end{eqnarray}
where the $G(N,\sigma)$ factors are truncated Gaussian densities
constraints using the estimated numbers of events and the associated
uncertainties.  We assume a uniform prior for $\sigma\cdot {\cal B}$
and integrate the likelihood over all parameters except
$\sigma\cdot {\cal B}$.  A 95\% credibility level upper limit on
$\sigma\cdot{\cal B}$ is obtained by calculating the $95^{th}$
percentile of the resulting distributions.

To measure the expected sensitivity for this analysis, background-only
pseudo-experiments are used to calculate an expected limit in the
absence of Higgs boson production.  Pseudo-data are generated by
fluctuating the individual background estimates within total
uncertainties.  The expected limit is derived from the pseudo-data
using Eq.~\ref{eq:comboLike}.

The likelihoods from events with exactly one {\sc secvtx} $b$-tagged jet passing
the NN $b$-tagging filter and events with at least two
{\sc secvtx} $b$-tagged jets criteria are multiplied together.
The systematic uncertainties associated with the pretag acceptance,
 luminosity uncertainty, and uncertainty of the $b$-tagging efficiency
 scale factor are considered to be 100\% correlated between the two
 selection channels.  Background uncertainties, specifically on the
 heavy-flavor fractions and $b$-tagging scale factor, are also
 completely correlated.
The ``=1 tag w/ NNtag'' selection combined with ``$\geq$2
 Tag'' gives the best expected limit, as expected from the sensitivity
 study (see Fig.~\ref{fig:sensitivity}).

\begin{figure}
  \begin{center}
   \includegraphics[width=0.48\textwidth]{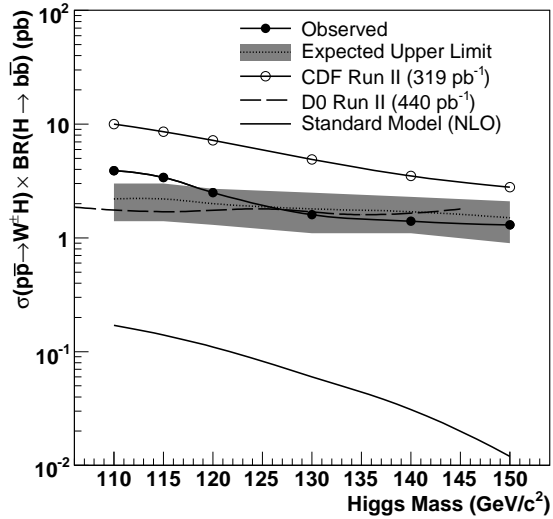}
   \caption{95\% confidence level upper limit on
     $\sigma(p\bar{p}\rightarrow WH)\cdot {\cal B}(H\rightarrow
     b\bar{b})$ with an integrated luminosity of $1$~fb$^{-1}$
     obtained from the combined likelihood between events with exactly
     one {\sc secvtx} $b$-tag passing the NN b-tagging and events
     with at least two {\sc secvtx} $b$-tagged jets.  The previous CDF
     data~\cite{Abulencia:2006aj} and recent D0
     data~\cite{new_d0} are shown for comparison.}
   \label{fig:upperLimit_sdnntag}
  \end{center}
 \end{figure}

\begin{figure}[htbp]
  \begin{center}
    \includegraphics*[width=0.40\textwidth]{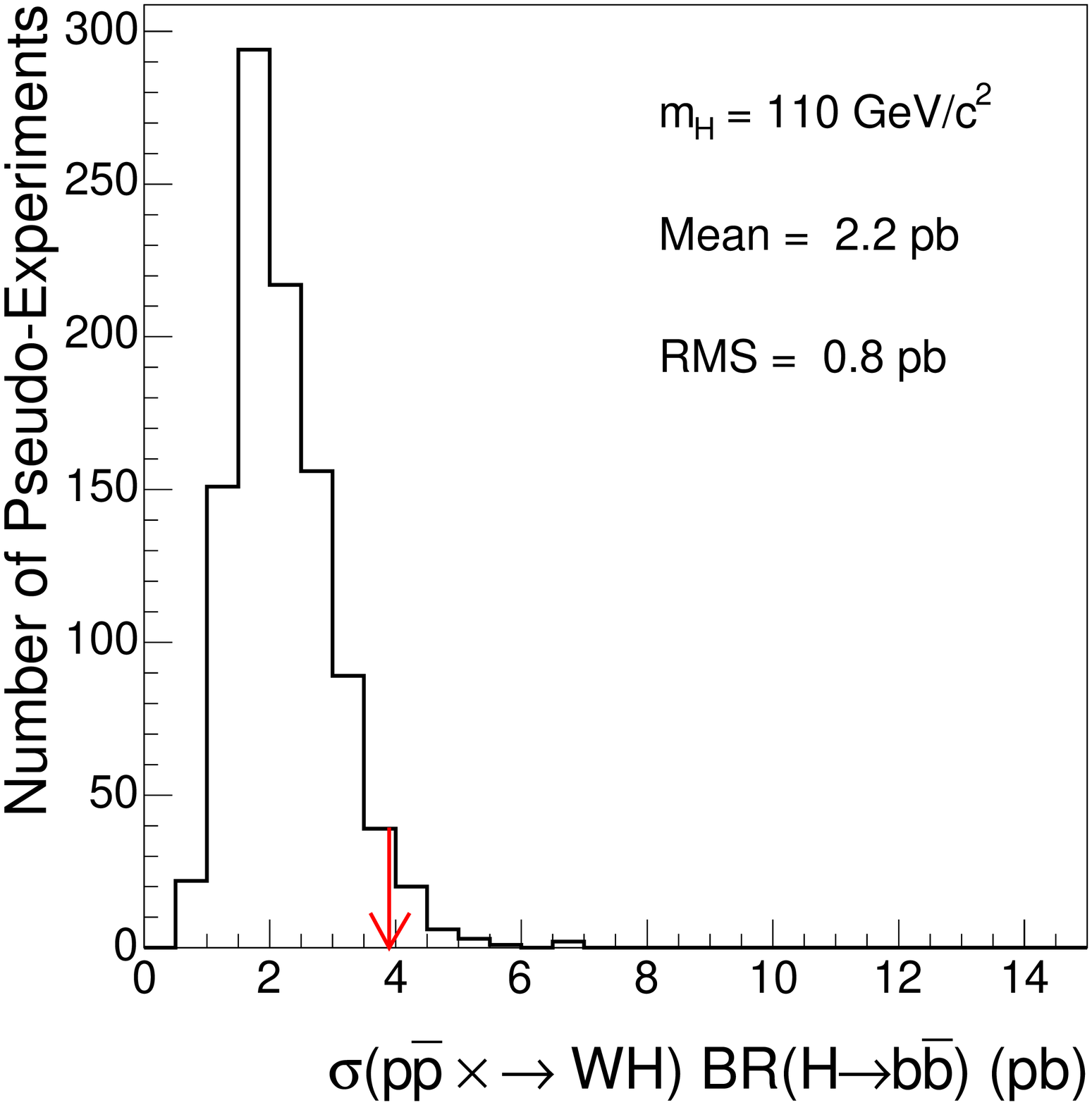}
    \includegraphics*[width=0.40\textwidth]{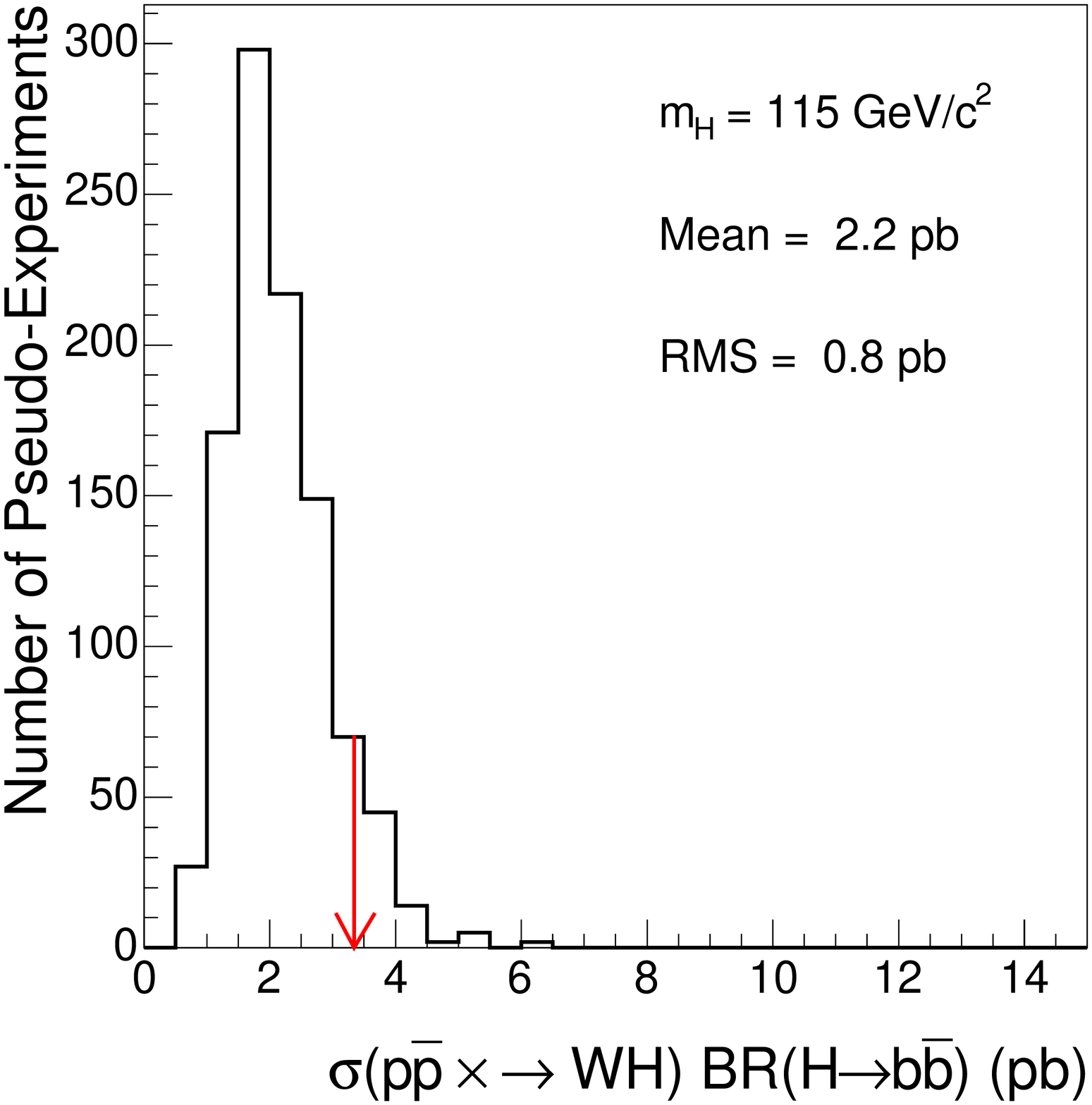}
    \includegraphics*[width=0.40\textwidth]{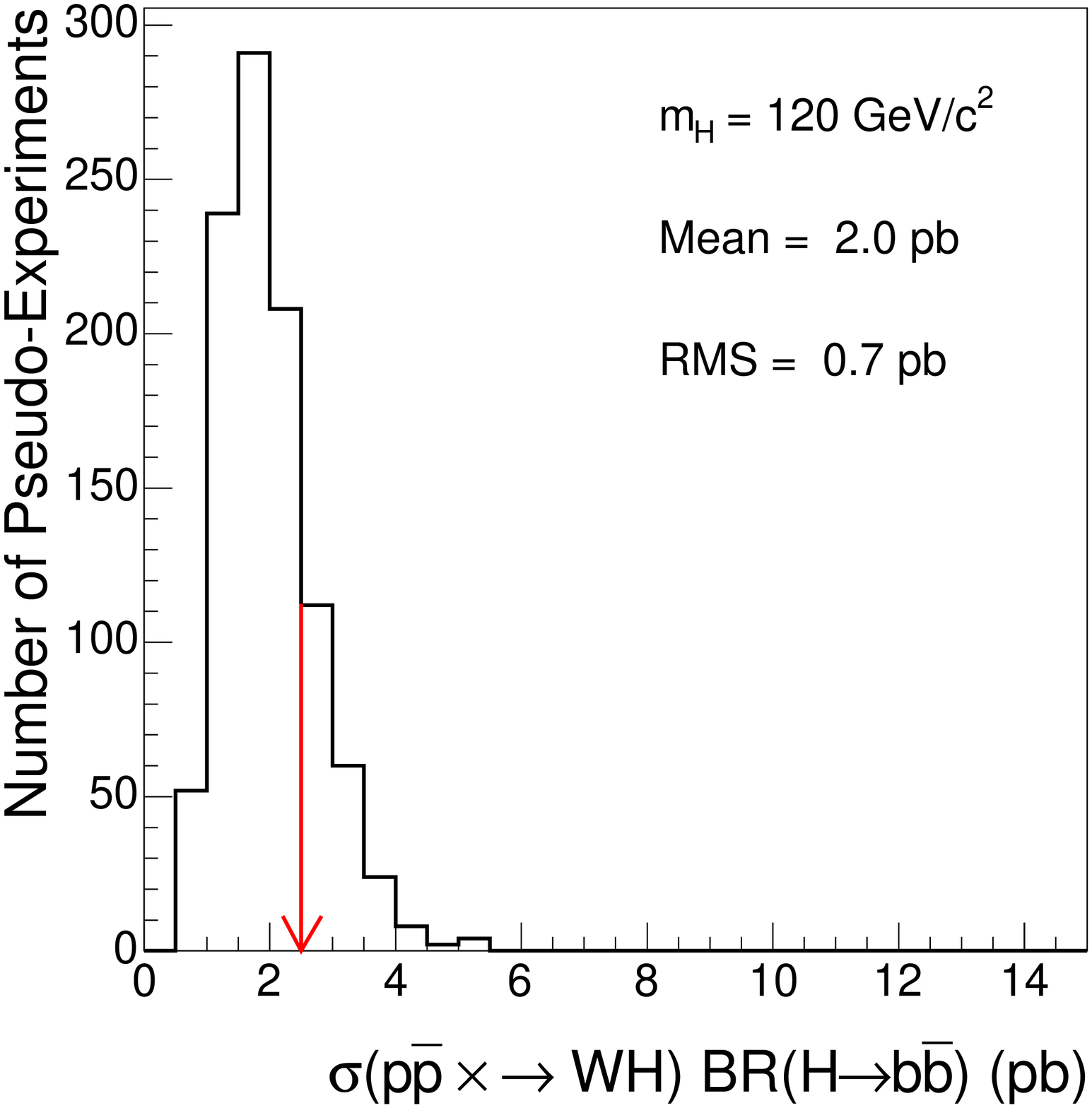}
    \includegraphics*[width=0.40\textwidth]{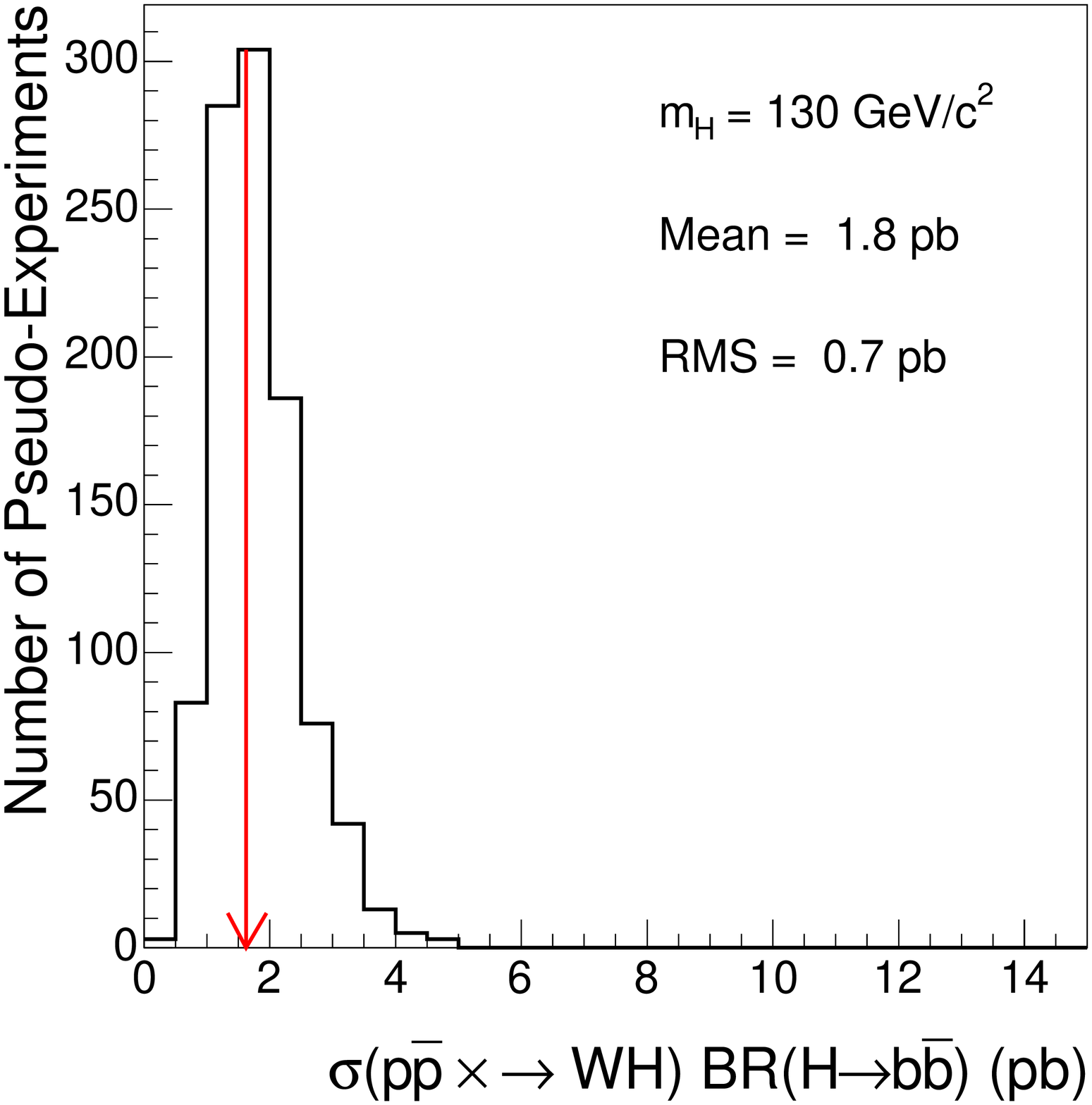}
    \includegraphics*[width=0.40\textwidth]{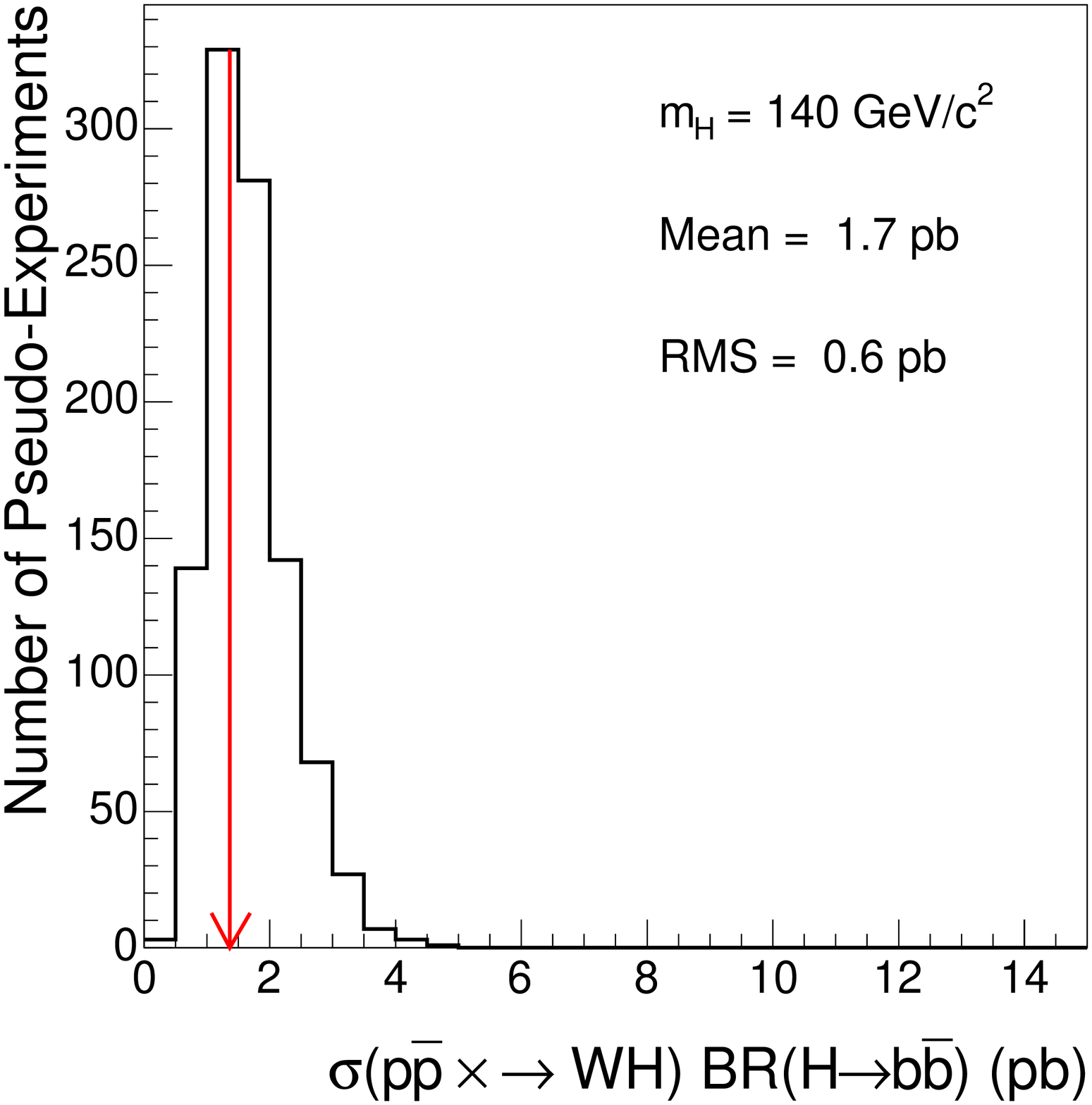}
    \includegraphics*[width=0.40\textwidth]{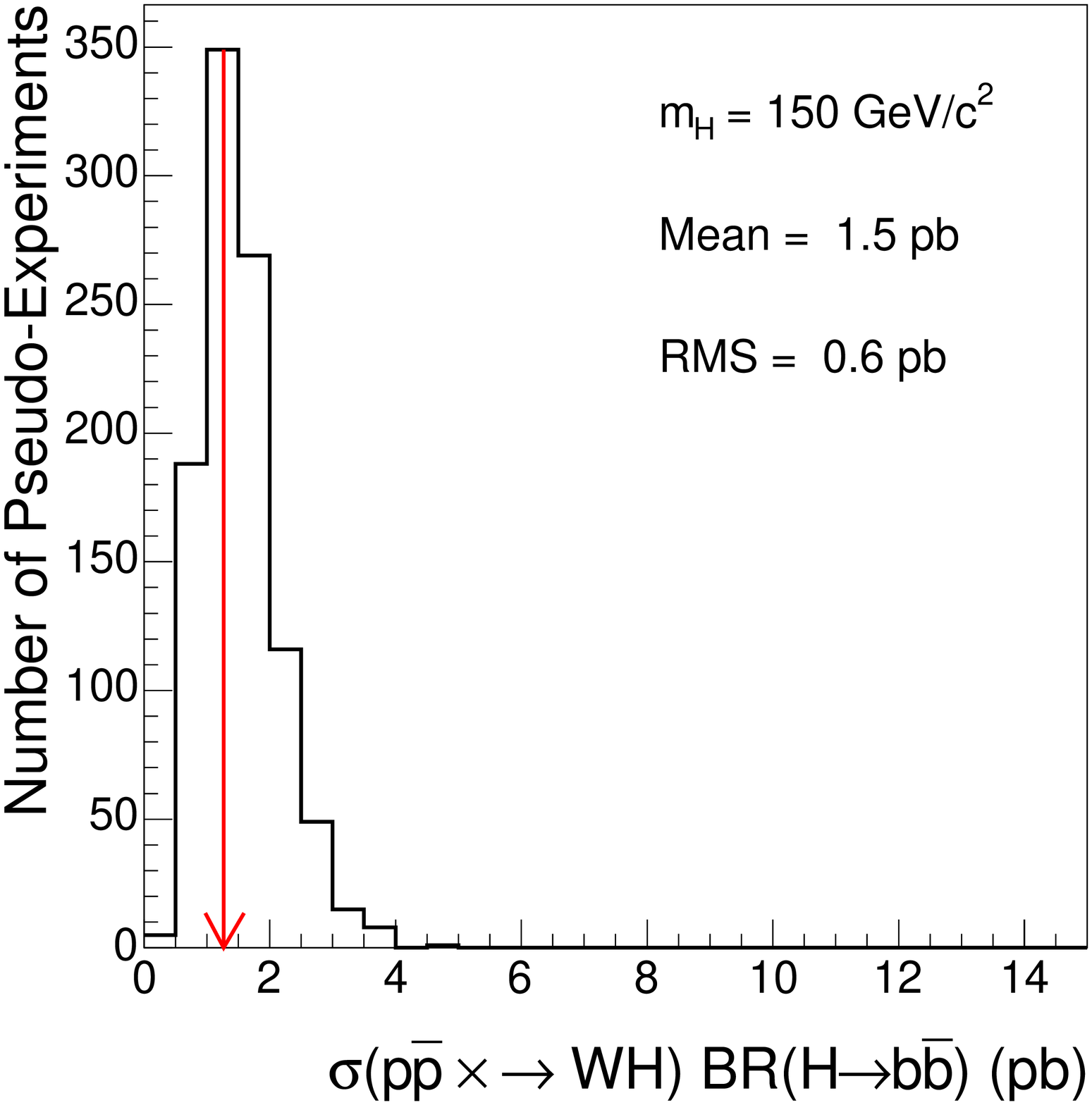}
    \caption{Results of 95\% confidence level limits obtained from the
  combined likelihood in pseudo-experiments. The arrows indicate the observed limits.
  \label{fig:pseudoExps_sdnntag}}
  \end{center}
\end{figure}

The observed limits as a function of the Higgs boson
 mass are shown in Fig.~\ref{fig:upperLimit_sdnntag} and
 Table~\ref{tbl:upperLimit_sdnntag}, together with the expected limits
 determined from pseudo-experiments. An ensemble of limits from
 pseudo-experiments and the observed limit for each Higgs boson mass
 point are shown in Fig.~\ref{fig:pseudoExps_sdnntag}.  The limit in
 the low mass region is at most two standard deviations higher than the
 expected limit, but this is consistent with a statistical
 fluctuation in the dijet mass distributions (see
 Fig.~\ref{fig:m2j_snntag}) around $m_H=115\,\mathrm{GeV}/c^2$.  Such a
 fluctuation is much larger than the expectation for SM
 Higgs boson production in this channel.

The search sensitivity is improved significantly with respect to
previous searches, about 30\% beyond the expectations from simple luminosity
scaling.  The two main effects are the separation of the $b$-tagged
data sample into single- and double-tagged events, and the NN filter
applied to the single-tag sample.

\begin{table}
  \begin{center}
    \begin{tabular}{cccc}
      \hline
      \hline
      Higgs Mass    &     \multicolumn{2}{c}{Upper Limit (pb)} \\
       GeV/c$^{2}$  &      Observed  & Expected      & SM \\
      \hline
      110  &               3.9       &  2.2$\pm$0.8   & 0.16     \\
      115  &               3.4       &  2.2$\pm$0.8   & 0.13     \\
      120  &               2.5       &  2.0$\pm$0.7   & 0.10     \\
      130  &               1.6       &  1.8$\pm$0.7   & 0.060     \\
      140  &               1.4       &  1.7$\pm$0.6   & 0.030     \\
      150  &               1.3       &  1.5$\pm$0.6   & 0.011     \\
      \hline
    \end{tabular}
   \caption{Observed and expected upper limits on $\sigma(p\bar{p}\rightarrow WH)\cdot {\cal B}(H\rightarrow b\bar{b})$ at 95~\%~C.L., compared to the SM production rate calculated at NNLO.}
    \label{tbl:upperLimit_sdnntag}
  \end{center}
\end{table}

\section{Conclusions}
\label{sec:conclusions}
We have presented a search for the standard model Higgs boson in the
$\ell \nu b\bar b$ final state expected from $WH$ production.  The
event selection includes an additional neural network $b$-tag filter
to reduce the background contributions from light flavor
and charm quark jets.  This improvement, along with a total dataset
corresponding to $1\,\mathrm{fb}^{-1}$, allows us to improve the upper
limit on Higgs boson production.  We set a 95\% confidence level upper
limit on the production cross section times branching ratio varying
from 3.9 to 1.3~pb for Higgs boson masses 110 to
$150\,\mathrm{GeV}/c^2$.

\begin{acknowledgments}

We thank the Fermilab staff and the technical staffs of the
participating institutions for their vital contributions. This work
was supported by the U.S. Department of Energy and National Science
Foundation; the Italian Istituto Nazionale di Fisica Nucleare; the
Ministry of Education, Culture, Sports, Science and Technology of
Japan; the Natural Sciences and Engineering Research Council of
Canada; the National Science Council of the Republic of China; the
Swiss National Science Foundation; the A.P. Sloan Foundation; the
Bundesministerium f\"ur Bildung und Forschung, Germany; the Korean
Science and Engineering Foundation and the Korean Research Foundation;
the Science and Technology Facilities Council and the Royal Society,
UK; the Institut National de Physique Nucleaire et Physique des
Particules/CNRS; the Russian Foundation for Basic Research; the
Comisi\'on Interministerial de Ciencia y Tecnolog\'{\i}a, Spain; the
European Community's Human Potential Programme; the Slovak R\&D Agency;
and the Academy of Finland.

\end{acknowledgments}


\input{prd_wh1invfb.bbl}
\end{document}

%% file: Jan_2008_Authors_Visitors1.tex
\affiliation{Institute of Physics, Academia Sinica, Taipei, Taiwan 11529, Republic of China} 
\affiliation{Argonne National Laboratory, Argonne, Illinois 60439} 
\affiliation{University of Athens, 157 71 Athens, Greece} 
\affiliation{Institut de Fisica d'Altes Energies, Universitat Autonoma de Barcelona, E-08193, Bellaterra (Barcelona), Spain} 
\affiliation{Baylor University, Waco, Texas  76798} 
\affiliation{Istituto Nazionale di Fisica Nucleare Bologna, $^t$University of Bologna, I-40127 Bologna, Italy} 
\affiliation{Brandeis University, Waltham, Massachusetts 02254} 
\affiliation{University of California, Davis, Davis, California  95616} 
\affiliation{University of California, Los Angeles, Los Angeles, California  90024} 
\affiliation{University of California, San Diego, La Jolla, California  92093} 
\affiliation{University of California, Santa Barbara, Santa Barbara, California 93106} 
\affiliation{Instituto de Fisica de Cantabria, CSIC-University of Cantabria, 39005 Santander, Spain} 
\affiliation{Carnegie Mellon University, Pittsburgh, PA  15213} 
\affiliation{Enrico Fermi Institute, University of Chicago, Chicago, Illinois 60637} 
\affiliation{Comenius University, 842 48 Bratislava, Slovakia; Institute of Experimental Physics, 040 01 Kosice, Slovakia} 
\affiliation{Joint Institute for Nuclear Research, RU-141980 Dubna, Russia} 
\affiliation{Duke University, Durham, North Carolina  27708} 
\affiliation{Fermi National Accelerator Laboratory, Batavia, Illinois 60510} 
\affiliation{University of Florida, Gainesville, Florida  32611} 
\affiliation{Laboratori Nazionali di Frascati, Istituto Nazionale di Fisica Nucleare, I-00044 Frascati, Italy} 
\affiliation{University of Geneva, CH-1211 Geneva 4, Switzerland} 
\affiliation{Glasgow University, Glasgow G12 8QQ, United Kingdom} 
\affiliation{Harvard University, Cambridge, Massachusetts 02138} 
\affiliation{Division of High Energy Physics, Department of Physics, University of Helsinki and Helsinki Institute of Physics, FIN-00014, Helsinki, Finland} 
\affiliation{University of Illinois, Urbana, Illinois 61801} 
\affiliation{The Johns Hopkins University, Baltimore, Maryland 21218} 
\affiliation{Institut f\"{u}r Experimentelle Kernphysik, Universit\"{a}t Karlsruhe, 76128 Karlsruhe, Germany} 
\affiliation{Center for High Energy Physics: Kyungpook National University, Daegu 702-701, Korea; Seoul National University, Seoul 151-742, Korea; Sungkyunkwan University, Suwon 440-746, Korea; Korea Institute of Science and Technology Information, Daejeon, 305-806, Korea; Chonnam National University, Gwangju, 500-757, Korea} 
\affiliation{Ernest Orlando Lawrence Berkeley National Laboratory, Berkeley, California 94720} 
\affiliation{University of Liverpool, Liverpool L69 7ZE, United Kingdom} 
\affiliation{University College London, London WC1E 6BT, United Kingdom} 
\affiliation{Centro de Investigaciones Energeticas Medioambientales y Tecnologicas, E-28040 Madrid, Spain} 
\affiliation{Massachusetts Institute of Technology, Cambridge, Massachusetts  02139} 
\affiliation{Institute of Particle Physics: McGill University, Montr\'{e}al, Canada H3A~2T8; and University of Toronto, Toronto, Canada M5S~1A7} 
\affiliation{University of Michigan, Ann Arbor, Michigan 48109} 
\affiliation{Michigan State University, East Lansing, Michigan  48824} 
\affiliation{University of New Mexico, Albuquerque, New Mexico 87131} 
\affiliation{Northwestern University, Evanston, Illinois  60208} 
\affiliation{The Ohio State University, Columbus, Ohio  43210} 
\affiliation{Okayama University, Okayama 700-8530, Japan} 
\affiliation{Osaka City University, Osaka 588, Japan} 
\affiliation{University of Oxford, Oxford OX1 3RH, United Kingdom} 
\affiliation{Istituto Nazionale di Fisica Nucleare, Sezione di Padova-Trento, $^u$University of Padova, I-35131 Padova, Italy} 
\affiliation{LPNHE, Universite Pierre et Marie Curie/IN2P3-CNRS, UMR7585, Paris, F-75252 France} 
\affiliation{University of Pennsylvania, Philadelphia, Pennsylvania 19104} 
\affiliation{Istituto Nazionale di Fisica Nucleare Pisa, $^q$University of Pisa, $^r$University of Siena and $^s$Scuola Normale Superiore, I-56127 Pisa, Italy} 
\affiliation{University of Pittsburgh, Pittsburgh, Pennsylvania 15260} 
\affiliation{Purdue University, West Lafayette, Indiana 47907} 
\affiliation{University of Rochester, Rochester, New York 14627} 
\affiliation{The Rockefeller University, New York, New York 10021} 

\affiliation{Istituto Nazionale di Fisica Nucleare, Sezione di Roma 1, $^v$Sapienza Universit\`{a} di Roma, I-00185 Roma, Italy} 

\affiliation{Rutgers University, Piscataway, New Jersey 08855} 
\affiliation{Texas A\&M University, College Station, Texas 77843} 
\affiliation{Istituto Nazionale di Fisica Nucleare Trieste/\ Udine, $^w$University of Trieste/\ Udine, Italy} 
\affiliation{University of Tsukuba, Tsukuba, Ibaraki 305, Japan} 
\affiliation{Tufts University, Medford, Massachusetts 02155} 
\affiliation{Waseda University, Tokyo 169, Japan} 
\affiliation{Wayne State University, Detroit, Michigan  48201} 
\affiliation{University of Wisconsin, Madison, Wisconsin 53706} 
\affiliation{Yale University, New Haven, Connecticut 06520} 
\author{T.~Aaltonen}
\affiliation{Division of High Energy Physics, Department of Physics, University of Helsinki and Helsinki Institute of Physics, FIN-00014, Helsinki, Finland}
\author{J.~Adelman}
\affiliation{Enrico Fermi Institute, University of Chicago, Chicago, Illinois 60637}
\author{T.~Akimoto}
\affiliation{University of Tsukuba, Tsukuba, Ibaraki 305, Japan}
\author{M.G.~Albrow}
\affiliation{Fermi National Accelerator Laboratory, Batavia, Illinois 60510}
\author{B.~\'{A}lvarez~Gonz\'{a}lez}
\affiliation{Instituto de Fisica de Cantabria, CSIC-University of Cantabria, 39005 Santander, Spain}
\author{S.~Amerio$^u$}
\affiliation{Istituto Nazionale di Fisica Nucleare, Sezione di Padova-Trento, $^u$University of Padova, I-35131 Padova, Italy} 

\author{D.~Amidei}
\affiliation{University of Michigan, Ann Arbor, Michigan 48109}
\author{A.~Anastassov}
\affiliation{Northwestern University, Evanston, Illinois  60208}
\author{A.~Annovi}
\affiliation{Laboratori Nazionali di Frascati, Istituto Nazionale di Fisica Nucleare, I-00044 Frascati, Italy}
\author{J.~Antos}
\affiliation{Comenius University, 842 48 Bratislava, Slovakia; Institute of Experimental Physics, 040 01 Kosice, Slovakia}
% ...

\collaboration{CDF Collaboration\footnote{With visitors from $^a$Universiteit Antwerpen, B-2610 Antwerp, Belgium, 
$^b$Chinese Academy of Sciences, Beijing 100864, China, 
$^c$University of Bristol, Bristol BS8 1TL, United Kingdom, 
$^d$University of California Irvine, Irvine, CA  92697, 
$^e$University of California Santa Cruz, Santa Cruz, CA  95064, 
$^f$Cornell University, Ithaca, NY  14853, 
$^g$University of Cyprus, Nicosia CY-1678, Cyprus, 
$^h$University College Dublin, Dublin 4, Ireland, 
$^i$University of Edinburgh, Edinburgh EH9 3JZ, United Kingdom, 
$^j$Universidad Iberoamericana, Mexico D.F., Mexico, 
$^k$University of Manchester, Manchester M13 9PL, England, 
$^l$Nagasaki Institute of Applied Science, Nagasaki, Japan, 
$^m$University de Oviedo, E-33007 Oviedo, Spain, 
$^n$Queen Mary, University of London, London, E1 4NS, England, 
$^o$Texas Tech University, Lubbock, TX  79409, 
$^p$IFIC(CSIC-Universitat de Valencia), 46071 Valencia, Spain, 
}}
\noaffiliation

%% file: prd_wh1invfb.bbl
\begin{thebibliography}{31}
\expandafter\ifx\csname natexlab\endcsname\relax\def\natexlab#1{#1}\fi
\expandafter\ifx\csname bibnamefont\endcsname\relax
  \def\bibnamefont#1{#1}\fi
\expandafter\ifx\csname bibfnamefont\endcsname\relax
  \def\bibfnamefont#1{#1}\fi
\expandafter\ifx\csname citenamefont\endcsname\relax
  \def\citenamefont#1{#1}\fi
\expandafter\ifx\csname url\endcsname\relax
  \def\url#1{\texttt{#1}}\fi
\expandafter\ifx\csname urlprefix\endcsname\relax\def\urlprefix{URL }\fi
\providecommand{\bibinfo}[2]{#2}
\providecommand{\eprint}[2][]{\url{#2}}

\bibitem[{\citenamefont{Higgs}(1964)}]{Higgs:1964pj}
\bibinfo{author}{\bibfnamefont{P.~W.} \bibnamefont{Higgs}},
  \bibinfo{journal}{Phys. Rev. Lett.} \textbf{\bibinfo{volume}{13}},
  \bibinfo{pages}{508} (\bibinfo{year}{1964}).

\bibitem[{\citenamefont{Barate et~al.}(2003)}]{Barate:2003sz}
\bibinfo{author}{\bibfnamefont{R.}~\bibnamefont{Barate}} \bibnamefont{et~al.}
  (\bibinfo{collaboration}{ALEPH, DELPHI, L3, and OPAL collaborations and the
  LEP Working Group for Higgs boson searches}), \bibinfo{journal}{Phys. Lett.}
  \textbf{\bibinfo{volume}{B565}}, \bibinfo{pages}{61} (\bibinfo{year}{2003}),
  \eprint{hep-ex/0306033}.

\bibitem[{\citenamefont{Alcaraz et~al.}(2006)}]{Alcaraz:2006mx}
\bibinfo{author}{\bibfnamefont{J.}~\bibnamefont{Alcaraz}} \bibnamefont{et~al.}
  (\bibinfo{collaboration}{ALEPH, DELPHI, L3, and OPAL collaborations and the
  LEP Electroweak Working Group}), \bibinfo{type}{Tech. Rep.}
  \bibinfo{number}{CERN-PH-EP-2006-042} (\bibinfo{year}{2006}),
  \eprint{hep-ex/0612034}.

\bibitem[{\citenamefont{Han and Willenbrock}(1991)}]{Han:1991ia}
\bibinfo{author}{\bibfnamefont{T.}~\bibnamefont{Han}} \bibnamefont{and}
  \bibinfo{author}{\bibfnamefont{S.}~\bibnamefont{Willenbrock}},
  \bibinfo{journal}{Phys. Lett.} \textbf{\bibinfo{volume}{B273}},
  \bibinfo{pages}{167} (\bibinfo{year}{1991}).

\bibitem[{\citenamefont{Djouadi et~al.}(1998)\citenamefont{Djouadi, Kalinowski,
  and Spira}}]{Djouadi:1997yw}
\bibinfo{author}{\bibfnamefont{A.}~\bibnamefont{Djouadi}},
  \bibinfo{author}{\bibfnamefont{J.}~\bibnamefont{Kalinowski}},
  \bibnamefont{and} \bibinfo{author}{\bibfnamefont{M.}~\bibnamefont{Spira}},
  \bibinfo{journal}{Comput. Phys. Commun.} \textbf{\bibinfo{volume}{108}},
  \bibinfo{pages}{56} (\bibinfo{year}{1998}), \eprint{hep-ph/9704448}.

\bibitem[{\citenamefont{Abulencia et~al.}(2007)}]{Abulencia:2006ps}
\bibinfo{author}{\bibfnamefont{A.}~\bibnamefont{Abulencia}}
  \bibnamefont{et~al.} (\bibinfo{collaboration}{CDF collaboration}),
  \bibinfo{journal}{Phys. Rev.} \textbf{\bibinfo{volume}{D75}},
  \bibinfo{pages}{012010} (\bibinfo{year}{2007}), \eprint{hep-ex/0612015}.

\bibitem[{\citenamefont{Abulencia
  et~al.}(2006{\natexlab{a}})}]{Abulencia:2006aj}
\bibinfo{author}{\bibfnamefont{A.}~\bibnamefont{Abulencia}}
  \bibnamefont{et~al.} (\bibinfo{collaboration}{CDF collaboration}),
  \bibinfo{journal}{Phys. Rev. Lett.} \textbf{\bibinfo{volume}{97}},
  \bibinfo{pages}{081802} (\bibinfo{year}{2006}{\natexlab{a}}),
  \eprint{hep-ex/0605124}.

\bibitem[{\citenamefont{Abazov et~al.}(to be published)}]{new_d0}
\bibinfo{author}{\bibfnamefont{V.~M.} \bibnamefont{Abazov}}
  \bibnamefont{et~al.} (\bibinfo{collaboration}{D0 collaboration}),
  \bibinfo{journal}{Phys. Lett. B}  (\bibinfo{year}{to be published}),
  \eprint{hep-ex/0712.0598}.

\bibitem[{\citenamefont{Acosta et~al.}(2005{\natexlab{a}})}]{Acosta:2004hw}
\bibinfo{author}{\bibfnamefont{D.}~\bibnamefont{Acosta}} \bibnamefont{et~al.}
  (\bibinfo{collaboration}{CDF collaboration}), \bibinfo{journal}{Phys. Rev.}
  \textbf{\bibinfo{volume}{D71}}, \bibinfo{pages}{052003}
  (\bibinfo{year}{2005}{\natexlab{a}}), \eprint{hep-ex/0410041}.

\bibitem[{\citenamefont{Acosta et~al.}(2005{\natexlab{b}})}]{Acosta:2004yw}
\bibinfo{author}{\bibfnamefont{D.}~\bibnamefont{Acosta}} \bibnamefont{et~al.}
  (\bibinfo{collaboration}{CDF collaboration}), \bibinfo{journal}{Phys. Rev.}
  \textbf{\bibinfo{volume}{D71}}, \bibinfo{pages}{032001}
  (\bibinfo{year}{2005}{\natexlab{b}}), \eprint{hep-ex/0412071}.

\bibitem[{\citenamefont{Balka et~al.}(1988)}]{Balka:1987ty}
\bibinfo{author}{\bibfnamefont{L.}~\bibnamefont{Balka}} \bibnamefont{et~al.},
  \bibinfo{journal}{Nucl. Instrum. Methods} \textbf{\bibinfo{volume}{A267}},
  \bibinfo{pages}{272} (\bibinfo{year}{1988}).

\bibitem[{\citenamefont{Bertolucci et~al.}(1988)}]{Bertolucci:1987zn}
\bibinfo{author}{\bibfnamefont{S.}~\bibnamefont{Bertolucci}}
  \bibnamefont{et~al.}, \bibinfo{journal}{Nucl. Instrum. Methods}
  \textbf{\bibinfo{volume}{A267}}, \bibinfo{pages}{301} (\bibinfo{year}{1988}).

\bibitem[{\citenamefont{Albrow et~al.}(2002)}]{Albrow:2001jw}
\bibinfo{author}{\bibfnamefont{M.~G.} \bibnamefont{Albrow}}
  \bibnamefont{et~al.}, \bibinfo{journal}{Nucl. Instrum. Methods}
  \textbf{\bibinfo{volume}{A480}}, \bibinfo{pages}{524} (\bibinfo{year}{2002}).

\bibitem[{\citenamefont{Abe et~al.}(1992)}]{Abe:1991ui}
\bibinfo{author}{\bibfnamefont{F.}~\bibnamefont{Abe}} \bibnamefont{et~al.}
  (\bibinfo{collaboration}{CDF collaboration}), \bibinfo{journal}{Phys. Rev.}
  \textbf{\bibinfo{volume}{D45}}, \bibinfo{pages}{1448} (\bibinfo{year}{1992}).

\bibitem[{\citenamefont{Bhatti et~al.}(2006)}]{Bhatti:2005ai}
\bibinfo{author}{\bibfnamefont{A.}~\bibnamefont{Bhatti}} \bibnamefont{et~al.},
  \bibinfo{journal}{Nucl. Instrum. Methods} \textbf{\bibinfo{volume}{A566}},
  \bibinfo{pages}{375} (\bibinfo{year}{2006}), \eprint{hep-ex/0510047}.

\bibitem[{\citenamefont{Ascoli et~al.}(1988)}]{Ascoli:1987av}
\bibinfo{author}{\bibfnamefont{G.}~\bibnamefont{Ascoli}} \bibnamefont{et~al.},
  \bibinfo{journal}{Nucl. Instrum. Methods} \textbf{\bibinfo{volume}{A268}},
  \bibinfo{pages}{33} (\bibinfo{year}{1988}).

\bibitem[{\citenamefont{Dorigo}(2001)}]{Dorigo:2000ip}
\bibinfo{author}{\bibfnamefont{T.}~\bibnamefont{Dorigo}}
  (\bibinfo{collaboration}{CDF collaboration}), \bibinfo{journal}{Nucl.
  Instrum. Methods} \textbf{\bibinfo{volume}{A461}}, \bibinfo{pages}{560}
  (\bibinfo{year}{2001}).

\bibitem[{\citenamefont{Thomson et~al.}(2002)}]{Thomson:2002xp}
\bibinfo{author}{\bibfnamefont{E.~J.} \bibnamefont{Thomson}}
  \bibnamefont{et~al.}, \bibinfo{journal}{IEEE Trans. Nucl. Sci.}
  \textbf{\bibinfo{volume}{49}}, \bibinfo{pages}{1063} (\bibinfo{year}{2002}).

\bibitem[{\citenamefont{Acosta et~al.}(2005{\natexlab{c}})}]{Acosta:2004uq}
\bibinfo{author}{\bibfnamefont{D.}~\bibnamefont{Acosta}} \bibnamefont{et~al.}
  (\bibinfo{collaboration}{CDF collaboration}), \bibinfo{journal}{Phys. Rev.
  Lett.} \textbf{\bibinfo{volume}{94}}, \bibinfo{pages}{091803}
  (\bibinfo{year}{2005}{\natexlab{c}}), \eprint{hep-ex/0406078}.

\bibitem[{\citenamefont{Abulencia et~al.}(2006{\natexlab{b}})}]{Sal:1}
\bibinfo{author}{\bibfnamefont{A.}~\bibnamefont{Abulencia}}
  \bibnamefont{et~al.} (\bibinfo{collaboration}{CDF collaboration}),
  \bibinfo{journal}{Phys. Rev. Lett.} \textbf{\bibinfo{volume}{97}},
  \bibinfo{pages}{082004} (\bibinfo{year}{2006}{\natexlab{b}}).

\bibitem[{\citenamefont{Peterson et~al.}(1994)\citenamefont{Peterson,
  R{\"{o}}gnvaldsson, and L{\"{o}}nnblad}}]{Peterson:1993nk}
\bibinfo{author}{\bibfnamefont{C.}~\bibnamefont{Peterson}},
  \bibinfo{author}{\bibfnamefont{T.}~\bibnamefont{R{\"{o}}gnvaldsson}},
  \bibnamefont{and}
  \bibinfo{author}{\bibfnamefont{L.}~\bibnamefont{L{\"{o}}nnblad}},
  \bibinfo{journal}{Comput. Phys. Commun.} \textbf{\bibinfo{volume}{81}},
  \bibinfo{pages}{185} (\bibinfo{year}{1994}).

\bibitem[{\citenamefont{Abulencia
  et~al.}(2006{\natexlab{c}})}]{Abulencia:2006kv}
\bibinfo{author}{\bibfnamefont{A.}~\bibnamefont{Abulencia}}
  \bibnamefont{et~al.} (\bibinfo{collaboration}{CDF collaboration}),
  \bibinfo{journal}{Phys. Rev.} \textbf{\bibinfo{volume}{D74}},
  \bibinfo{pages}{072006} (\bibinfo{year}{2006}{\natexlab{c}}),
  \eprint{hep-ex/0607035}.

\bibitem[{\citenamefont{Abulencia
  et~al.}(2006{\natexlab{d}})}]{Abulencia:2006in}
\bibinfo{author}{\bibfnamefont{A.}~\bibnamefont{Abulencia}}
  \bibnamefont{et~al.} (\bibinfo{collaboration}{CDF collaboration}),
  \bibinfo{journal}{Phys. Rev. Lett.} \textbf{\bibinfo{volume}{97}},
  \bibinfo{pages}{082004} (\bibinfo{year}{2006}{\natexlab{d}}),
  \eprint{hep-ex/0606017}.

\bibitem[{\citenamefont{Mangano et~al.}(2003)\citenamefont{Mangano, Moretti,
  Piccinini, Pittau, and Polosa}}]{Mangano:2002ea}
\bibinfo{author}{\bibfnamefont{M.~L.} \bibnamefont{Mangano}},
  \bibinfo{author}{\bibfnamefont{M.}~\bibnamefont{Moretti}},
  \bibinfo{author}{\bibfnamefont{F.}~\bibnamefont{Piccinini}},
  \bibinfo{author}{\bibfnamefont{R.}~\bibnamefont{Pittau}}, \bibnamefont{and}
  \bibinfo{author}{\bibfnamefont{A.~D.} \bibnamefont{Polosa}},
  \bibinfo{journal}{J. High Energy Phys.} \textbf{\bibinfo{volume}{07}},
  \bibinfo{pages}{001} (\bibinfo{year}{2003}), \eprint{hep-ph/0206293}.

\bibitem[{\citenamefont{Corcella et~al.}(2001)}]{Corcella:2001wc}
\bibinfo{author}{\bibfnamefont{G.}~\bibnamefont{Corcella}} \bibnamefont{et~al.}
  (\bibinfo{year}{2001}), \eprint{hep-ph/0201201}.

\bibitem[{\citenamefont{Campbell and Ellis}(2002)}]{Campbell:2002tg}
\bibinfo{author}{\bibfnamefont{J.}~\bibnamefont{Campbell}} \bibnamefont{and}
  \bibinfo{author}{\bibfnamefont{R.~K.} \bibnamefont{Ellis}},
  \bibinfo{journal}{Phys. Rev.} \textbf{\bibinfo{volume}{D65}},
  \bibinfo{pages}{113007} (\bibinfo{year}{2002}), \eprint{hep-ph/0202176}.

\bibitem[{\citenamefont{Cacciari et~al.}(2004)\citenamefont{Cacciari, Frixione,
  Mangano, Nason, and Ridolfi}}]{Cacciari:2003fi}
\bibinfo{author}{\bibfnamefont{M.}~\bibnamefont{Cacciari}},
  \bibinfo{author}{\bibfnamefont{S.}~\bibnamefont{Frixione}},
  \bibinfo{author}{\bibfnamefont{M.~L.} \bibnamefont{Mangano}},
  \bibinfo{author}{\bibfnamefont{P.}~\bibnamefont{Nason}}, \bibnamefont{and}
  \bibinfo{author}{\bibfnamefont{G.}~\bibnamefont{Ridolfi}},
  \bibinfo{journal}{J. High Energy Phys.} \textbf{\bibinfo{volume}{04}},
  \bibinfo{pages}{068} (\bibinfo{year}{2004}), \eprint{hep-ph/0303085}.

\bibitem[{\citenamefont{Harris et~al.}(2002)\citenamefont{Harris, Laenen, Phaf,
  Sullivan, and Weinzierl}}]{Harris:2002md}
\bibinfo{author}{\bibfnamefont{B.~W.} \bibnamefont{Harris}},
  \bibinfo{author}{\bibfnamefont{E.}~\bibnamefont{Laenen}},
  \bibinfo{author}{\bibfnamefont{L.}~\bibnamefont{Phaf}},
  \bibinfo{author}{\bibfnamefont{Z.}~\bibnamefont{Sullivan}}, \bibnamefont{and}
  \bibinfo{author}{\bibfnamefont{S.}~\bibnamefont{Weinzierl}},
  \bibinfo{journal}{Phys. Rev.} \textbf{\bibinfo{volume}{D66}},
  \bibinfo{pages}{054024} (\bibinfo{year}{2002}), \eprint{hep-ph/0207055}.

\bibitem[{\citenamefont{Sj{\"{o}}strand et~al.}(2001)}]{Sjostrand:2000wi}
\bibinfo{author}{\bibfnamefont{T.}~\bibnamefont{Sj{\"{o}}strand}}
  \bibnamefont{et~al.}, \bibinfo{journal}{Comput. Phys. Commun.}
  \textbf{\bibinfo{volume}{135}}, \bibinfo{pages}{238} (\bibinfo{year}{2001}),
  \eprint{hep-ph/0010017}.

\bibitem[{\citenamefont{Abulencia
  et~al.}(2006{\natexlab{e}})}]{Abulencia:2005aj}
\bibinfo{author}{\bibfnamefont{A.}~\bibnamefont{Abulencia}}
  \bibnamefont{et~al.} (\bibinfo{collaboration}{CDF collaboration}),
  \bibinfo{journal}{Phys. Rev.} \textbf{\bibinfo{volume}{D73}},
  \bibinfo{pages}{032003} (\bibinfo{year}{2006}{\natexlab{e}}),
  \eprint{hep-ex/0510048}.

\bibitem[{\citenamefont{Pumplin et~al.}(2002)}]{Pumplin:2002vw}
\bibinfo{author}{\bibfnamefont{J.}~\bibnamefont{Pumplin}} \bibnamefont{et~al.},
  \bibinfo{journal}{J. High Energy Phys.} \textbf{\bibinfo{volume}{07}},
  \bibinfo{pages}{012} (\bibinfo{year}{2002}), \eprint{hep-ph/0201195}.

\end{thebibliography}
